\documentclass[sigconf]{acmart}

\usepackage{multirow}
\usepackage{float}
\usepackage{stfloats}
\usepackage{url}
\usepackage{makecell}
\usepackage{xcolor}
\definecolor{mycolor}{HTML}{00FF00}

\AtBeginDocument{%
  }

\copyrightyear{2025} 
\acmYear{2025} 
\setcopyright{cc} 
\setcctype{by-nc}
\acmConference[CHI '25]{CHI Conference on Human Factors in Computing Systems}{April 26-May 1, 2025}{Yokohama, Japan}
\acmBooktitle{CHI Conference on Human Factors in Computing Systems (CHI '25), April 26-May 1, 2025, Yokohama, Japan}\acmDOI{10.1145/3706598.3713770}
\acmISBN{979-8-4007-1394-1/25/04}

\begin{document}

\title{Understanding How Psychological Distance Influences User Preferences in Conversational versus Web Search}


\author{Yitian Yang}
\orcid{0009-0000-7530-2116}
\affiliation{%
  \department{Computer Science}
  \institution{National University of Singapore}
  \city{Singapore}
  \country{Singapore}
}
\email{yang.yitian@u.nus.edu}

\author{Yugin Tan}
\orcid{0009-0006-7357-0436}
\affiliation{%
  \department{Computer Science}
  \institution{National University of Singapore}
  \city{Singapore}
  \country{Singapore}
}
\email{tan.yugin1@gmail.com}

\author{Yang Chen Lin}
\orcid{0000-0002-2477-4110}
\affiliation{%
  \department{Institute of Neuroscience}
  \institution{National Yang Ming Chiao Tung University}
  \city{Taipei}
  \country{Taiwan}
}
\email{yangchenlin@nctu.edu.tw}

\author{Jung-Tai King}
\orcid{0000-0003-2745-5999}
\affiliation{%
  \department{Hua-Shih College of Education}
  \institution{National Dong Hwa University}
  \city{Hualien}
  \country{Taiwan}
}
\email{jtchin2@gmail.com}

\author{Zihan Liu}
\orcid{0009-0003-5956-7132}
\affiliation{%
  \department{School of Information Sciences}
  \institution{University of Illinois Urbana-Champaign}
  \city{Champaign}
  \state{Illinois}
  \country{USA}
}
\email{zihanl18@illinois.edu}

\author{Yi-Chieh Lee}
\orcid{0000-0002-5484-6066}
\affiliation{%
  \department{Computer Science}
  \institution{National University of Singapore}
  \city{Singapore}
  \country{Singapore}
}
\email{yclee@nus.edu.sg}


\begin{abstract}
Conversational search offers an easier and faster alternative to conventional web search, while having downsides like a lack of source verification. Research has examined performance disparities between these two systems in various settings. However, little work has investigated how changes in the nature of a search task affect user preferences. We investigate how psychological distance - the perceived closeness of one to an event - affects user preferences between conversational and web search. We hypothesise that tasks with different psychological distances elicit different information needs, which in turn affect user preferences between systems. Our study finds that, under fixed condition ordering, greater psychological distances lead users to prefer conversational search, which they perceive as more credible, useful, enjoyable, and easy to use. We reveal qualitative reasons for these differences and provide design implications for search system designers.
\end{abstract}








\maketitle
\section{Introduction}

As digital technologies continue to advance, the ways people interact with information have significantly evolved. Users are no longer confined to traditional web search engines like Google Search \cite{kobayashi2000information, brin2012reprint}, but increasingly have the option of conversational search interfaces powered by Large Language Models (LLMs) like ChatGPT \cite{ai2023information, liao2020conversational}. Conversational search systems can potentially increase usability \cite{tariverdiyeva2019chatbots}, enhance efficiency \cite{ai2023information} by synthesising information from extensive underlying datasets \cite{deng2022benefits}, and increase user enjoyment through humanlike dialogues \cite{rhim2022application, folstad2020users, brandtzaeg2017people}. However, such systems also have certain downsides: hallucination of facts and information \cite{ji2023survey} and lack of source attribution can reduce perceptions of information completeness and verifiability \cite{jung2024we}, negatively affecting user trust, acceptance, and system adoption. Prior work has investigated user preferences for these two information retrieval (IR) systems in a range of tasks. Conversational search was found to provide superior information to web search in informal medical diagnosis \cite{sandmann2024systematic, van2024if, shen2024chatgpt}, and lower cognitive load \cite{kumar2024understanding} and greater efficacy \cite{arias2023comparing} in coding-related queries.

However, a crucial overlooked factor is the \textit{inherent variation in individual tasks}. Tasks can vary significantly in terms of the perceived \textit{psychological distance} to the user: for instance, whether they are seeking information for themselves or someone else, or whether they are searching about an event more or less likely to happen. The well-established Construal Level Theory (CLT) describes the relation between psychological distance and the level of abstraction of thought \cite{trope2010construal}. The psychological distance of a task (e.g. planning a trip for next week versus next year) influences whether people think about specific details (e.g. public transport options) or general ideas (e.g. cities to visit) \cite{trope2010construal}.
Such variations significantly impact how users process information and make decisions \cite{lermer2015effect, trope2010construal, wakslak2009effect, fiedler2007construal}; we propose that this effect extends to information needs in IR tasks, and thus user perceptions and preferences of the IR systems that fulfill them. Psychological distance is a common factor in everyday life, being assessed automatically without conscious effort \cite{liberman2014traversing}, and has fundamental and widespread importance for cognitive and behavioral decision making \cite{liberman2014traversing, maglio2013common}. Therefore, the effect of psychological distance of IR tasks on preferences for IR systems presents a key research gap.

To investigate this, we conducted a 2x4 between-subjects study, with 128 participants divided into eight groups. Each participant was assigned to \textbf{one of four psychological distance dimensions} posited by CLT — spatial (where an event takes place), temporal (when an event occurs), social (who is involved in the event), or hypothetical (how likely the event is to happen) — and \textbf{one of two psychological distance levels} (near or far). Participants completed an IR task, manipulated according to their assigned dimension and distance, first using web search and then conversational search. They were then asked to rate the ease of use, enjoyment, and perceived credibility of each tool, and finally their preference between tools.

Our findings indicated that participants in the \textbf{far psychological distance} condition exhibited a \textbf{significantly higher level of preference for the LLM tool} over traditional web search, reporting greater ease of use, usefulness, enjoyment, and credibility. Additionally, participants' search behaviors varied depending on psychological distance. When distance was perceived as greater, participants more effectively utilized the strengths of the LLM system to explore broad ideas, such as cities, cuisines, and accommodations. In contrast, nearer distances prompted participants to focus on concrete steps for planning and executing their trips. These findings suggest that the choice of IR system is closely tied to the user's situational context, and offers important design implications for designers of IR systems.
 
Our work makes the following contributions: 
\begin{enumerate}
\item It advances our understanding of the differences in user perceptions and preferences when interacting with information through traditional search engines and LLM-powered conversational search, contributing to the literature on Human-Information Interaction. Additionally, we explore how these preferences can inform tool recommendations based on psychological distance for information seeking. 
\item It offers valuable insights for optimizing future IR systems to better align their functional designs with varying user needs across diverse contexts, ultimately enhancing user experience and satisfaction in an increasingly complex digital landscape. Notably, our findings suggest that conversational search systems should be prioritized for tasks involving greater psychological distance, as this factor significantly influences user acceptance of IR technologies. 
\end{enumerate}
\section{Related Work}

\subsection{Conversational Search as a New Means of IR for Information Seeking}
\label{sec:isir}

Information seeking originates from the library sciences, but has been increasingly studied in the context of computer science \cite{blandford2022interacting}. Information seeking activities comprise three main processes: experiencing information \textit{needs}, information \textit{seeking}, and information \textit{use} \cite{blandford2022interacting, choo2013web}. The former two are relevant to our current study. Information needs arise from problems encountered in specific situations and are influenced by multiple factors, such as individual knowledge gaps \cite{allen1996information}, the level of information need \cite{taylor1968question}, and the constraints of time and resources \cite{choo2000information}. Information seeking involves "identifying, selecting, and interacting with information sources to purposely retrieve information" \cite{choo2013web}. Each specific information seeking task influences both information needs, which in turn affects an individual's choice of information sources while looking for information \cite{choo2000information}, and information seeking, where the nature of the task affects the amount of effort and psychological state invested in the task \cite{cole2011theory, bystrom1995task}. Users gravitate toward information sources that are most relevant to their needs \cite{case2016looking}, and the personal sense of control that comes from finding information that matches their needs encourages further use of its source \cite{wilson1997information}. In sum, the nature of an information seeking task affects a user's choice of information source, and hence their choice of information seeking tool. This motivates our study of how the nature of tasks affects preferences for different IR systems.

IR is the task of searching for relevant information from an online repository to fulfill a user's need. 
Conversational search, a specialized branch of IR systems, aims to help users obtain information that meets their needs through flexible and natural dialogues, based on their queries and relevant feedback \cite{kiesel2021meta, jannach2021survey, liao2020conversational, radlinski2017theoretical, croft1987i3r}. 
LLMs provide a revolutionary step forward in empowering such systems and their advantages due to their outstanding performance in understanding and generating human-like text \cite{gao2020recent}.

Conversational search offers a number of innate advantages over traditional web search. Unlike web search queries that require specific terms and keywords, conversational search enables users to express their needs through natural dialogue, reducing the effort needed to clarify their particular aims \cite{radlinski2017theoretical} and reducing cognitive load \cite{kaushik2019dialogue}. For example, \cite{kaushik2023comparing} found that users perceived less cognitive load and greater usability with a conversational search interface, compared to a standard search interface with the same knowledge base. For search tasks that involve multiple query aspects, such as making travel plans that include accommodation, food, and attractions, conversational search can be preferable in allowing users to assess the overall utility of such multi-item compositions \cite{radlinski2017theoretical}. Furthermore, the natural conversational abilities and human-like qualities of LLMs provide greater social interaction, which itself is valuable in enhancing the overall search experience \cite{folstad2020users}.

However, LLM-powered conversational search also suffers from a few disadvantages. LLMs, as natural language generation models \cite{brown2020language}, inevitably experience hallucination \cite{maynez2020faithfulness}, generating information that contradicts or is not supported by their source content \cite{ji2023survey}. Moreover, LLMs often do not provide specific sources for the summaries they generate \cite{bender2021dangers}. Finally, because their training data are based on past internet information with a specific cut-off point \cite{achiam2023gpt}, the information provided by LLMs can be outdated, and they may be unable to answer queries that require the latest data or information \cite{liu2023summary}. This therefore raises the question of when LLMs may or may not be superior to traditional web search. We address this issue in this study by asking participants to directly compare the two search systems in a realistic use case.

\subsection{Construal Level Theory and Psychological Distance}
\label{sec:clt}

To explore how specific tasks influence user preferences for web or conversational search, we turn to CLT, which provides a theoretical framework for manipulating task situation through psychological distance. Psychological distance is a common and critical factor in information seeking tasks, such as planning for near or distant events, or searching information for oneself or someone else. These variations in psychological distance can significantly impact how users process information and make decisions \cite{lermer2015effect, trope2010construal, wakslak2009effect, fiedler2007construal}. CLT not only provides ways to manipulate the psychological distance of tasks, but also provides a theoretical basis for the potential impact on user preferences towards different IR systems. When users make choices or set preferences for different objects, they rely on their understanding of these objects \cite{trope2010construal}. In CLT, increasing psychological distance enhances higher-level understanding, which in turn influences user evaluations and preferences \cite{trope2010construal}. CLT posits that as the psychological distance of a mental phenomenon increases, the construal level of thought processes shifts from low-level concrete constructions to high-level abstract constructions \cite{trope2010construal}. 

CLT suggests that as distance and hence construal levels shift, differing mental representations of a given task affect how people approach that task \cite{trope2010construal} - specifically, they affect the information and features that people find salient and relevant to task completion. For instance, distance was found to influence purchase-making decisions \cite{trope2000temporal}, with closer distances causing participants to prioritize central features of the product over its peripheral features. Similar effects were found in evaluation of student essays \cite{liviatan2008effect} and advice given for choosing a job \cite{kray1999differential}. We propose that psychological distance also affects the way users approach information seeking tasks. Since tasks with different psychological distances cause users to focus on obtaining different kinds of information, and web and conversational search fulfill different information needs, varying the psychological distances of tasks could cause differing preferences between the two search systems.

CLT proposes four main dimensions of psychological distance: spatial, temporal, social, and hypothetical \cite{trope2010construal}. Spatial distance is \textit{where} events are going to happen; temporal distance concerns \textit{when} events are going to happen; social distance is how \textit{similar} the people involved in an event are to oneself; and hypothetical distance is the \textit{likelihood} of an event happening \cite{trope2010construal}. Each dimension has some unique characteristics, but all four influence the mental models of a user through their construal level \cite{liberman2014traversing}. Low-level constructions are unstructured and detail-rich, encompassing primary, secondary, and incidental characteristics of an event \cite{trope2010construal}. In contrast, high-level constructions are more abstract and less specific, extracting essential points from available information while omitting details \cite{trope2010construal}. We propose that web search aligns more with the need for low-level constructions, while LLM-powered conversational search meets higher-level construal needs. Therefore, as the psychological distance associated with an information seeking task influences user information needs and information seeking behaviors through different construal levels, this in turn affects their perceptions and preferences for these two technologies. We explore this potential novel difference by presenting participants with tasks of different psychological distances, and comparing the resulting differences in user preference.

\subsection{Psychological Distance in Information Seeking Tasks}

Previous research has highlighted the potential advantages of conversational search in terms of ease of use (reducing user effort), usefulness (especially for complex tasks), and enjoyability, as well as potential issues with credibility. 
For user preferences, the effects of these interactive aspects are certainly important. In addition, the nature of the information seeking tasks is also an aspect worth considering. As mentioned in section \ref{sec:isir}, the nature of information seeking tasks influences users' choices of information sources, which should affect their preferences for different IR systems.
Previous studies have explored the relative efficacy of conversational search and web search across various types of information seeking tasks, particularly in domains such as healthcare and tourism. In the healthcare context, Hristidis et al. \cite{hristidis2023chatgpt} demonstrated that ChatGPT provides more relevant and objective responses to dementia-related queries compared to Google, though it lacks source reliability and timestamped information. Similarly, Shahsavar et al. \cite{shahsavar2023user} identified that users are increasingly inclined to use ChatGPT for self-diagnosis due to its higher performance expectancy and positive risk-reward appraisals, despite concerns about its suitability for healthcare purposes. Sandmann et al. \cite{sandmann2024systematic} further highlighted that GPT-4 outperforms Google in diagnostic accuracy across diverse clinical cases, showcasing the growing potential of LLMs in medical question answering. 
In the tourism domain, Zarezadeh et al. \cite{zarezadeh2023online} emphasized the potential of AI tools like ChatGPT to streamline information searches. Wong et al. \cite{wong2023autonomous} illustrated ChatGPT's ability to enhance decision-making during different stages of travel by providing personalized and cost-effective recommendations. Zhang et al. \cite{zhang2024enhancing} extended this by showcasing ChatGPT 4.0's multimodal capabilities, enabling users to compare destinations and plan activities. Finally, Dwivedi et al. \cite{dwivedi2024leveraging} outlined the transformative potential of generative AI in the hospitality and tourism industry while identifying challenges related to bias and data privacy.
Despite these advancements, these studies also highlight the limitations in conversational search, closely related to the challenges associated with LLMs mentioned in section \ref{sec:isir}, underscoring the importance of traditional web search \cite{sandmann2024systematic}. Users frequently turn to web search to verify LLM-generated information, indicating that LLMs complement rather than replace web search in information seeking tasks \cite{yen2024search}.

However, even if the type of task remains constant, psychological distance, a common factor in information seeking tasks, can also lead to changes in user information needs. Few studies have explored the impact of psychological distance in information seeking tasks. Moreover, previous studies have rarely focused on user perceptions and preferences for these two IR systems, lacking a comprehensive perspective from task nature to technology preference.
Therefore, this study aims to fill this gap by considering the psychological distance in information seeking tasks to deepen our understanding of user preferences for conversational search versus web search in real-world information searching. This not only helps us to understand the motivations behind user choices at specific task psychological distances, but also provides important insights for designing IR systems that better meet user needs.
\section{Current Study}

\begin{figure*}[ht]
    \centering
    \includegraphics[width=0.8\textwidth]{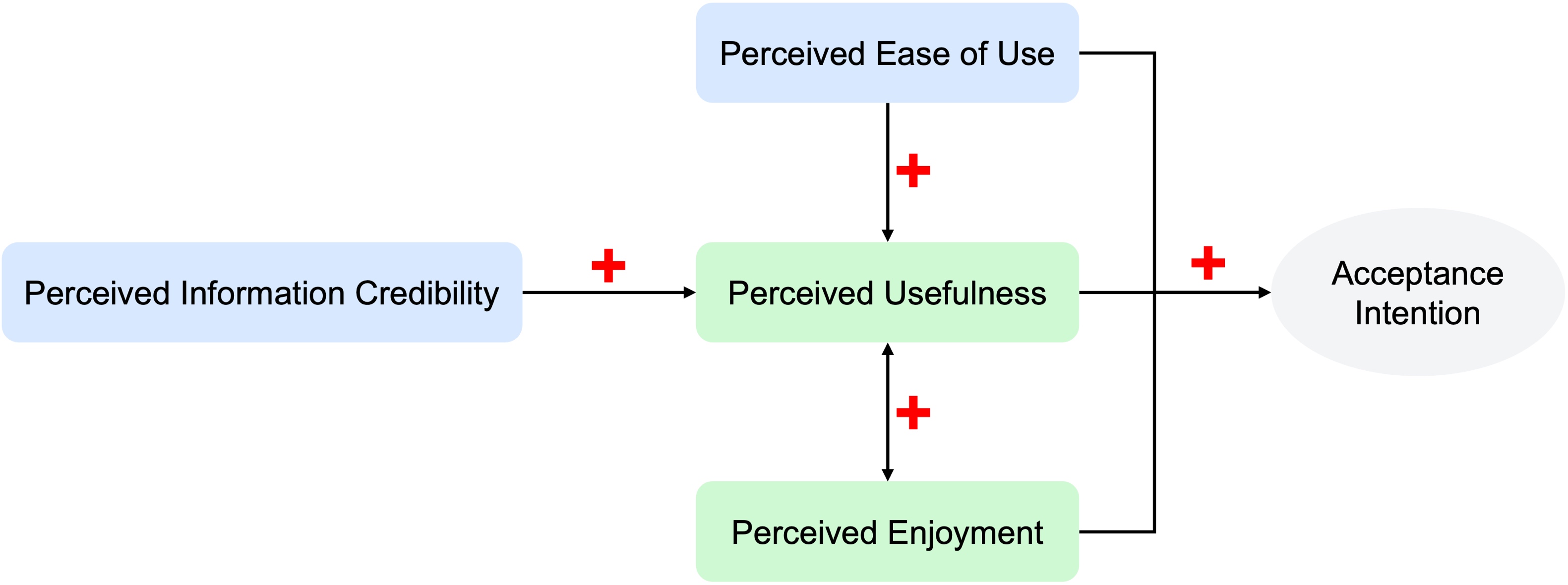}
    \caption{Extended TAMs incorporating perceived enjoyment and information credibility. This model illustrates the relationships between perceived ease of use, perceived usefulness, perceived enjoyment, and information credibility with the acceptance intention of technology.}
    \Description{This diagram visually represents an extended Technology Acceptance Model (TAM). It consists of four main rectangular components, each labeled with a different factor influencing technology acceptance. These components are connected by arrows indicating the direction of influence, each marked with a plus sign to denote positive relationships. Starting from the left, the first component is labeled "Perceived Information Credibility", which connects directly to "Perceived Usefulness", a central component in the middle of the diagram. Above "Perceived Usefulness" is "Perceived Ease of Use", which influences "Perceived Usefulness". Below "Perceived Usefulness" is "Perceived Enjoyment", linked to it by a bidirectional arrow, indicating a mutual positive influence between the two. "Perceived Ease of Use", "Perceived Usefulness" and "Perceived Enjoyment" all have direct connections to "Acceptance Intention", depicted as an oval on the far right, indicating the final outcome of the model. The diagram uses a color scheme where "Perceived Usefulness" and "Perceived Enjoyment" are highlighted in light green, while "Perceived Information Credibility" and "Perceived Ease of Use" are in light blue, representing a special focus on "Perceived Information Credibility" and "Perceived Ease of Use".}
    \label{fig:tam}
\end{figure*}

As mentioned in section \ref{sec:clt}, CLT provides a theoretical foundation for both manipulating tasks through the critical factor (psychological distances) and understanding their impact on user preferences. We investigate this gap by manipulating the psychological distance of information seeking tasks using CLT. We are particularly interested in exploring:

\begin{itemize}
    \item \textbf{RQ}: How psychological distance of information seeking tasks influences user preferences for web search or LLM-powered conversational search.
\end{itemize}

Web search conventionally provides a list of results ranked according to the relevance of the user's query, allowing users to access specific information through links \cite{brin1998anatomy}. However, obtaining a comprehensive overview of a topic is more difficult as users must access and filter through multiple sources and aggregate this information themselves. Conversely, LLM-powered conversational search generates summaries of the information relevant to the user's current search intent, providing a more comprehensive overview of relevant information. However, these tend to lack specific details \cite{achiam2023gpt, brown2020language, devlin2018bert}, and require users to ask further questions or refine their prompts for more detailed inquiries. Further, when users require highly-specific or time-sensitive information, LLM-powered conversational search may not provide accurate answers (e.g. local train ticket prices or current weather conditions).

CLT posits that tasks with greater psychological distances elicit preferences for more abstract, higher-levels overviews of information. Thus, we propose that the perceived ease of use of each IR system depends on the psychological distance of the task and thus the needs of the user. We hypothesise:

\begin{itemize}
    \item \textbf{H1}: Users engaging in information seeking tasks with greater psychological distances perceive greater ease of use with LLM-powered conversational search compared to web search, relative to users engaging in tasks with smaller psychological distances.
\end{itemize}

CLT suggests that as psychological distance increases, users tend to seek more supportive rather than opposing viewpoints, due to high-level thinking's tendency to overlook specific disadvantages.
In conversational search, recent studies \cite{sharma2024generative} suggest that LLM-powered conversational search creates an echo chamber effect, producing content that validates user views more than web search and reducing the occurrences of opposing viewpoints. Content that corroborates user views is perceived as more credible \cite{nickerson1998confirmation}, specifically in terms of accuracy, authenticity, and believability \cite{appelman2016measuring}. Therefore, we hypothesise the following.

\begin{itemize}
    \item \textbf{H2}: Users engaging in information seeking tasks with greater psychological distances perceive higher informational credibility in LLM-powered conversational search compared to web search, relative to users engaging in tasks with smaller psychological distances.
\end{itemize}

Technology Acceptance Models (TAMs) \cite{davis1989perceived, davis1992extrinsic, ayeh2015travellers} offer a theoretical basis for exploring the user acceptance and other related perceptions of conversational search by examining user perceptions of the ease of use and information credibility.
TAM posits that two primary factors influence a user's decision to adopt a technology: perceived usefulness and perceived ease of use \cite{davis1989perceived}. These perceptions directly influence users' attitudes towards the technology, while ease of use also indirectly affects attitudes by influencing perceived usefulness \cite{adams1992perceived}.

Extensions of the TAM have identified additional relevant variables beyond perceived usefulness and ease of use. One such factor is perceived enjoyment. Davis et al. \cite{davis1992extrinsic} expanded the TAM to include perceived enjoyment, which can also directly enhance user attitudes and perceived usefulness: when users find a technology enjoyable, they also perceive it as more useful, and vice versa. In another extension of the TAM for consumer-generated media that aids travel planning, Ayeh \cite{ayeh2015travellers} introduced information credibility as a prerequisite for perceived usefulness. 
In the context of IR, credibility plays an important role by indirectly influencing user preferences. Users are more likely to adopt and prefer systems that provide reliable information, as credibility directly affects their trust in the system \cite{mcknight2007factors}. When users trust the information provided by the system, they are more likely to perceive it as useful \cite{ayeh2015travellers, ayeh2013towards, mcknight2007factors}, which is consistent with the core structure of the TAM model \cite{davis1989perceived}.
Applying the TAM and its extensions in conversational search, we illustrate the interactions among user perceptions of these systems in Figure \ref{fig:tam}. Based on H1 and H2, we propose the following hypotheses:

\begin{itemize}
    \item \textbf{H3}: Users engaging in information seeking tasks with greater psychological distances perceive higher usefulness in LLM-powered conversational search compared to web search, relative to users engaging in tasks with smaller psychological distances.
    \item \textbf{H4}: Users engaging in information seeking tasks with greater psychological distances perceive higher enjoyment in LLM-powered conversational search compared to web search, relative to users engaging in tasks with smaller psychological distances.
    \item \textbf{H5}: Users engaging in information seeking tasks with greater psychological distances have higher acceptance in LLM-powered conversational search compared to web search, relative to users engaging in tasks with smaller psychological distances.
\end{itemize}
\section{Methodology}

Our experiment comprised eight groups, involving the four dimensions of psychological distance (spatial, temporal, social, and hypothetical) each with two levels of distance (near and far), constituting a 4x2 design. Participants in each experimental group used two IR technologies: a web search (WebSearch) system and an LLM-powered conversational search (ConvSearch) system.

\subsection{Task Design}

\begin{table*}
  \caption{Manipulations of Psychological Distance Across Four Dimensions in Information Seeking Tasks}
  \label{tab:distancemanipulations}
  \begin{tabular}{lcccccccc}
    \toprule
    \multicolumn{2}{l}{\textbf{Dimension}} & \textbf{Spatial} & & \textbf{Temporal} & & \textbf{Social} & & \textbf{Hypothetical} \\
    \midrule
    \multicolumn{2}{l}{\textbf{Manipulation}} & Trip destination & & Trip timing & & Traveller identity & & Likelihood of trip occurring \\
    \midrule
    \multicolumn{2}{l}{\textbf{Psych. Dist.}} & \multicolumn{7}{c}{\textbf{Task Description}} \\
    \midrule
    \multicolumn{2}{l}{Near} & Nearby country & & Next week & & Close friend of participant & & Confirmed, plane tickets booked \\
    \multicolumn{2}{l}{Far} & Faraway country & & Next year & & Internet user on online forum & & Not confirmed, only a possibility \\
  \bottomrule
\end{tabular}
\end{table*}

Participants were asked to complete an information seeking task with fictional travel scenarios. The tasks required participants to imagine planning a one-week vacation to another country, focusing on finding food, natural attractions, and historical buildings destinations.

Each participant was randomly presented with one of eight varying tasks. Each task was either \textbf{near} or \textbf{far} in psychological distance to the participant, and psychological distance was manipulated in one of four dimensions (spatial, temporal, social, or hypothetical). For instance, a participant in the (near, spatial) condition was asked to plan a trip to a nearby country, while a participant in the (far, temporal) condition planned a trip in the distant future. This allowed us to explore whether there are differences in user perceptions and preferences towards IR technologies based on psychological distance, and whether these differences are consistent across the different dimensions or are influenced by other factors. Table \ref{tab:distancemanipulations} describes the ways in which all the different dimensions and distances were manipulated.\footnote{For each task, dimensions that were not varied were given a "neutral" setting. For example, in the (near, spatial) condition, the social, temporal, and hypothetical dimensions were kept neutral. These were to ensure that dimensions of psychological distance other than the one currently under focus remained neutral. In determining the neutral conditions, we aimed to prevent these other dimensions from influencing participants' perception of the psychological distance being studied. For example, the choice of a destination country near to the residence of our participants reflects the reality that most international travel typically involves neighboring countries \cite{mckercher2019impact, mckercher2008impact}. Similarly, in describing the traveler in dimensions other than social distance, we chose the participant themselves because conducting information searches and planning trips for oneself is more common in everyday life.} \footnote{Details on the specific scenarios are given in full in the Appendix (A.1), uploaded with Supplementary Material.}.

Each participant performed the same tasks but with different destinations for WebSearch and ConvSearch, so that the first search using WebSearch did not influence the subsequent search with ConvSearch. In each experimental group, the difference in destination did not affect psychological distance: Nearby countries in the set of near spatial distances were designated as Mexico (WebSearch task) and Cuba (ConvSearch task), both geographically close to our participants in the United States, while countries in the far spatial distance settings were represented by Egypt (WebSearch task) and Sudan (ConvSearch task). These countries were chosen to minimize the influence of prior knowledge by excluding mainstream American travel destinations and focusing on non-Western countries.

\subsection{Procedure}

\begin{figure*}[htbp]
    \centering
    \includegraphics[width=1\textwidth]{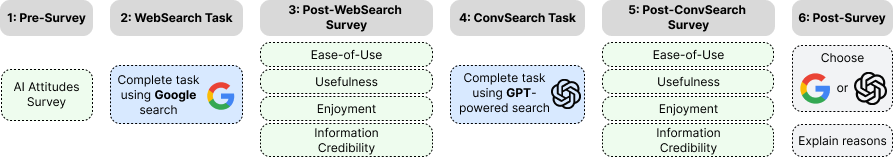}
    \caption{Flowchart of the study procedure detailing the sequence of tasks and surveys conducted across two information seeking tasks. This diagram illustrates the comprehensive steps from the initial pre-task survey through two distinct information seeking tasks, followed by post-task surveys and a final preference survey.}
    \Description{This figure presents a detailed flowchart of the study procedure, organized into six main phases across a horizontal layout. Each phase is represented as a rectangular box containing lists of activities or survey components, with arrows indicating the sequence from left to right. Step 1 is Pre Task Survey: This initial phase assesses participants' baseline attitudes towards AI. Step 2 is Information Seeking Task 1: This phase starts with reading a scenario, followed by using the WebSearch to gather information. Participants' search behaviors are recorded, and they are required to complete a travel plan. Step 3 is Post Task 1 Survey: This survey evaluates participants' perceptions after the first task, focusing on Perceived Ease of Use, Usefulness, Enjoyment, and Information Credibility. Step 4 is Information Seeking Task 2: Similar to the first task, participants read a new scenario and use a Conversational Search. This phase also involves recording the conversation and completing a travel plan. Step 5 is  Post Task 2 Survey: This survey mirrors the first post-task survey but evaluates the conversational search experience, assessing the same four perception metrics. Preference Survey: The final phase involves a survey where participants choose their preferred information retrieval system between WebSearch and the conversational search. They also provide reasons for their choice.}
    \label{fig:procedure}
\end{figure*}

The full study design (Figure \ref{fig:procedure}) includes six phases. In Step 1, participants completed a pre-task survey to measure their baseline attitudes to artificial intelligence (AI). In Step 2, the scenario for the first information seeking task was introduced. Participants then used WebSearch to seek information. We placed WebSearch first as it is a widely recognized IR system, and serves as a baseline for the experiment. To monitor the search behaviors of participants, we created a custom web search interface that integrated Google's search API \footnote{https://developers.google.com/custom-search/v1/overview}. After gathering the necessary information for the task, participants were required to write a travel plan consisting of at least 100 words to ensure the task completion quality. Step 3 was the first post-task survey, where participants were asked to rate their perceptions of ease of use, usefulness, enjoyment, and information credibility when using WebSearch. 
Steps 4 and 5 mirrored the previous two steps. Participants used ConvSearch to perform the second information seeking task, which had a different destination that had the same spatial distance as the first one. We built the ConvSearch system on the UChat\footnote{https://uchat.au/} platform using OpenAI's GPT-4 API \footnote{https://platform.openai.com/docs/api-reference/completions/object}, similar to ChatGPT. Participants used our tool to seek information and then reported their perceptions of this technology in the second post-task survey. To minimize cognitive load and ensure consistency in participants' experiences, we intentionally placed WebSearch first. It serves as a baseline for the experiment, as it is a system that participants are more familiar with in their daily lives. Previous studies have shown that using a system with which users are already familiar can reduce the cognitive load associated with the task \cite{blackler2010investigating, aarts1998predicting, norman1975data}, ensuring that users complete Steps 4 and 5 in a more natural state. Furthermore, as our study focuses on the relative comparisons between WebSearch and ConvSearch under different psychological distances, any potential bias introduced by the fixed order is consistent across all conditions and unlikely to confound the key findings. Finally, presenting the less familiar system first could have introduced additional variability, as participants might require more time to adapt to the new interface, which could inadvertently affect their perceptions \cite{limayenm2003habit}. Finally, in Step 6, participants indicated their preference between the two IR systems they used and elaborated on reasons for their choice.

\subsection{Participants}

Participants were recruited on the CloudResearch Connect platform\footnote{https://www.cloudresearch.com/products/connect-for-researchers/}, with residency restricted to the United States to ensure uniformity in their perception of spatial distances to designated travel destinations. After excluding individuals who failed attention checks, the study consisted of 128 participants according to the power analysis result (16 per group; 62 females, 64 males, 2 undisclosed). The average age of the participants was 37.6 (SD = 10.5), and 63\% of the participants had at least a bachelor's degree. The study was expected to last 30 minutes, with participants compensated at CloudResearch's recommended rate of \$10 per hour. This research was reviewed and approved by the National University of Singapore Departmental Ethics Review Committee.

\subsection{Measures}

\subsubsection{User perception, behavior and preferences}

The main dependent variables were based on the TAM \cite{davis1989perceived} and its variants \cite{davis1992extrinsic,featherman2003predicting, lee2009factors}. We aimed to understand how psychological distance influences user behavior and preferences for WebSearch and ConvSearch through affecting these user perception variables. Most perception variables involved several questions, each measured on a 7-point Likert scale ranging from "Strongly disagree" (1) to "Strongly agree" (7). The mean score reported by participants for all questions within each variable was used to represent the score of that variable.

\begin{itemize}
    \item \textbf{Perceived ease of use} and \textbf{Perceived usefulness} are gauged using questions from TAM \cite{davis1989perceived}. For ease of use, questions include, "\textit{Learning to operate web search/LLM-powered conversational search would be easy for me}" and "\textit{I would find it easy to get web search/LLM-powered conversational search to do what I want to do}" among a total of six questions. Similarly, perceived usefulness is measured with six questions, such as "\textit{Using web search/LLM-powered conversational search in my job would enable me to accomplish my tasks more quickly}" and "\textit{Using web search/LLM-powered conversational search would improve my job performance}".
    \item \textbf{Perceived enjoyment} is measured using three specific questions from a TAM variant \cite{davis1992extrinsic}, such as "\textit{I find using web search/LLM-powered conversational search enjoyable}" and "\textit{The actual process of using web search/LLM-powered conversational search is pleasant}".
    \item \textbf{Perceived information credibility} is represented through three specific dimensions: \textit{accuracy}, \textit{authenticity}, and \textit{believability} (following \cite{appelman2016measuring}). We provide participants with three sliders, each ranging from 1 to 100, to indicate their perceptions of information credibility across three dimensions. We then calculate the mean of these three values to represent the overall perceived credibility of the message.
    \item \textbf{User preferences} are determined by asking participants which tool they would choose for similar tasks in the future and by requiring them to explain their reasons for this choice.
\end{itemize}

\subsubsection{Control Variables}

We assessed participants' general attitudes toward AI as a control variable in our analysis. This measurement was taken before informing participants about the task.

\begin{itemize}
    \item \textbf{General attitudes towards AI} were measured by asking participants to respond to statements that measured their comfort with, perceptions of, and willingness to engage with AI technologies (derived from \cite{durndell2002computer} e.g., "\textit{AI is bringing us into a bright new era}" and "\textit{AI makes me uncomfortable because I do not understand it}"). Responses were recorded using a 7-point Likert scale, where 1 corresponds to "Strongly disagree" and 7 to "Strongly agree."
\end{itemize}

While attitudes toward AI are not the primary variables we observe, they could potentially influence participants' perceptions of ConvSearch. Biases or preconceived notions about AI might affect how individuals evaluate and utilize IR systems. For example, participants with a positive attitude toward AI may be more inclined to perceive these tools as useful and trustworthy. Therefore, we consider participants' attitudes toward AI as a control variable to analyze its linear relationship with our dependent variables. This approach ensures that our analysis accurately reflects the impact of the tools themselves, without being confounded by individual biases toward artificial intelligence.

\subsection{Data Analysis}

To mitigate the influence of control variables, we first analyzed the relationship between participants' overall willingness towards AI and the primary observed variables using the Pearson correlation coefficient \cite{pearson1895vii} \footnote{Specific data are presented in the Appendix (A.2), uploaded with Supplementary Material.}.

To test our hypotheses, we conducted paired sample t-tests \cite{student1908probable} on the dependent variables (Ease of Use, Information Credibility, Usefulness, and Enjoyment). This method allows us to compare participants' perceptions of the same variables after using both WebSearch and ConvSearch in the same experiment. Initially, we assessed the normality of the observed variables using the Shapiro-Wilk test \cite{shapiro1965analysis} \footnote{Detailed Shapiro-Wilk test results for each dependent variable and the type of paired samples t-test used are available in the Appendix (A.5), uploaded with Supplementary Material.}. For datasets that followed a normal distribution, we applied the student's paired sample t-test to evaluate differences between conditions. If the normality assumption was violated, we used the Wilcoxon signed-rank test \cite{wilcoxon1992individual} as a non-parametric alternative. 
To control the False Discovery Rate (FDR) across multiple tests, we applied the Benjamini-Hochberg procedure \cite{benjamini1995controlling} as a correction for multiple comparisons to the 32 paired sample t-tests conducted across the four dependent variables and psychological distance dimensions. Adjusted p-values are reported unless otherwise noted.
For the binary data on user preference, we employed the chi-square test \cite{pearson1900x} to examine changes in preference for ConvSearch under different psychological distances.

To further address our research questions, we conducted a thematic analysis \cite{braun2006using} of the reasons participants (n = 128) reported for their tool preferences, aiming to provide user-centered design recommendations for IR tools. Initially, two researchers independently performed semantic (what the participants said) and latent (researchers' interpretations) analyses to generate preliminary codes without predefined categories. The research team refined these codes through multiple discussions and developed themes using a bottom-up inductive approach. Subsequently, two researchers coded the reasons participants provided, potentially assigning multiple themes to each reason.
\section{Results}

\subsection{Psychological Distance Increases Differences in User Perceptions }

\subsubsection{The Difference in Perceived Ease of Use Increased (H1 Partially Supported)}

\begin{figure*}[htbp]
    \centering
    \includegraphics[width=0.9\textwidth]{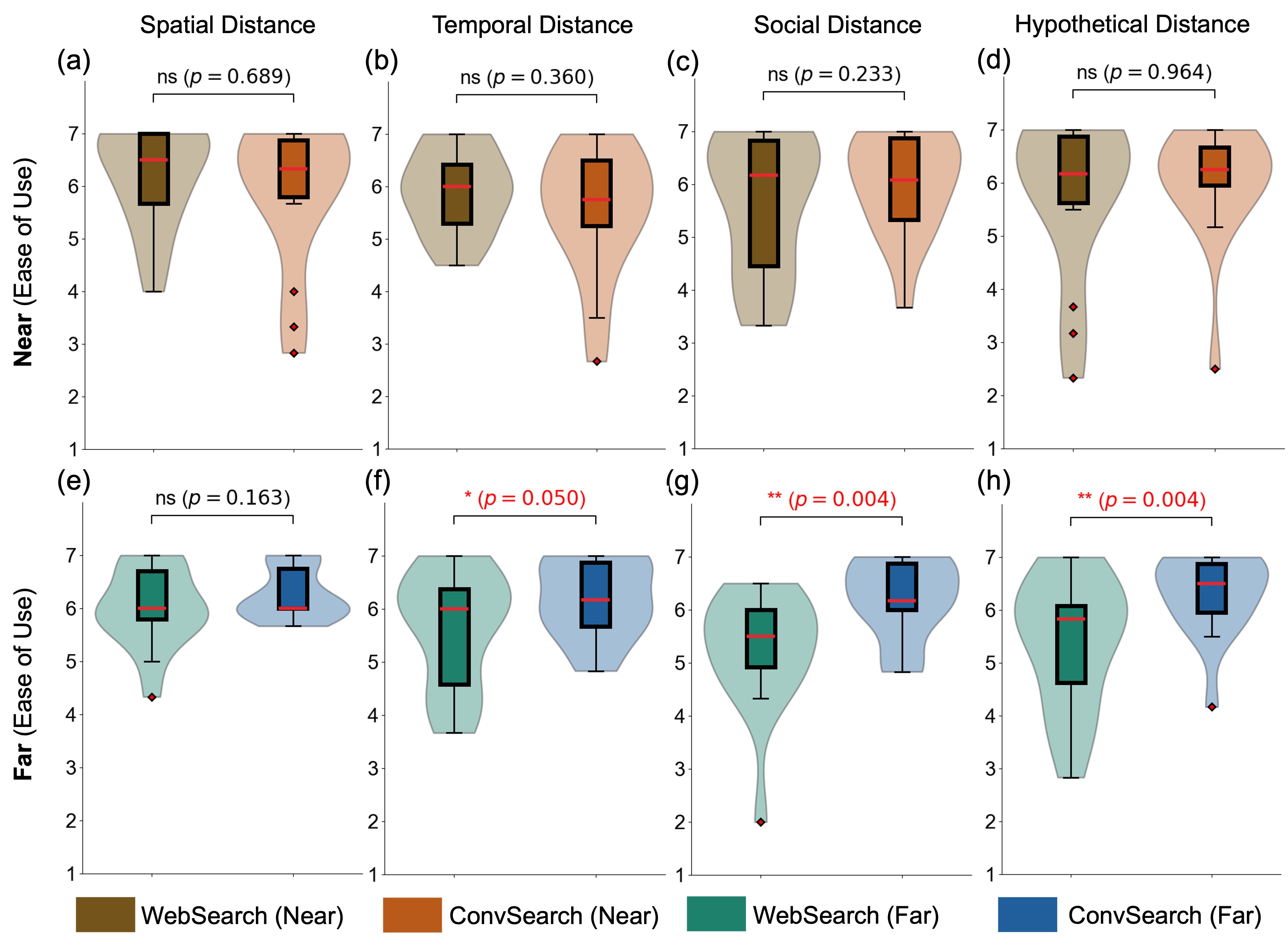}
    \caption{Comparison of perceived ease of use between WebSearch and ConvSearch in Far Psychological Distances. Panels (a)-(d) show results at a near psychological distance, and panels (e)-(h) at a far psychological distance, highlighting significant differences in the latter. Significance is marked as p > 0.05 (ns), p <= 0.05 (*), p < 0.01 (**), or p < 0.001 (***).}
    \Description{This figure presents eight violin plots arranged in two rows and four columns, comparing the perceived ease of use for two search technologies, WebSearch and ConvSearch, across four dimensions of psychological distance: spatial, temporal, social, and hypothetical. Each plot features a pair of violin plots side by side, with the left representing WebSearch and the right representing ConvSearch. The violin plots illustrate the distribution of responses, with a central box plot highlighting the median (red line), the interquartile range (colored box), and outliers (diamond shapes). Statistical significance is indicated with p-values and asterisks from the paired samples statistical tests. In the first row, panels (a) through (d) depict the perceived ease of use at a near psychological distance. The violin plots are colored in shades of beige and brown, representing WebSearch and ConvSearch respectively. The p-values are as follows: (a) Spatial Distance (p = 0.689), marked as non-significant (ns); (b) Temporal Distance (p = 0.360), marked as non-significant (ns); (c) Social Distance (p = 0.233), marked as non-significant (ns); (d) Hypothetical Distance (p = 0.964), marked as non-significant (ns). In the second row, panels (e) through (h) illustrate the perceived ease of use at a far psychological distance. The violin plots are colored in shades of green and blue, again representing WebSearch and ConvSearch respectively. Panel (e) Spatial Distance shows a p-value of (p = 0.163), marked as non-significant (ns). The p-values indicate significant differences in panels (f) Temporal Distance (p = 0.050), marked with one asterisk (*); (g) Social Distance (p = 0.004), marked with two asterisks (**); (h) Hypothetical Distance (p = 0.004), marked with two asterisks (**).}
    \label{fig:EoU}
\end{figure*}

When psychological distance was near, participants reported similar scores of perceived ease of use for WebSearch (Spatial Distance: \(M = 6.19\), \(SD = 0.96\); Temporal Distance: \(M = 5.91\), \(SD = 0.79\); Social Distance: \(M = 5.63\), \(SD = 1.32\); Hypothetical Distance: \(M = 5.81\), \(SD = 1.48\)) and ConvSearch (Spatial Distance: \(M = 5.91\), \(SD = 1.35\); Temporal Distance: \(M = 5.55\), \(SD = 1.22\); Social Distance: \(M = 6.01\), \(SD = 0.97\); Hypothetical Distance: \(M = 6.03\), \(SD = 1.09\)). As depicted in Fig.~\ref{fig:EoU} (a)-(d), paired samples statistical tests showed non-significant differences with low effect sizes (Spatial Distance: \(W = 32.50\), \(p = 0.689\), \(r = 0.18\); Temporal Distance: \(t(15) = 1.15\), \(p = 0.360\), Cohen's \(d = 0.29\); Social Distance: \(t(15) = -1.51\), \(p = 0.233\), Cohen's \(d = -0.38\); Hypothetical Distance: \(W = 32.00\), \(p = 0.964\), \(r = -0.03\)).  
The positive effect sizes for spatial and temporal distance suggest that when information seeking tasks are framed with near spatial or temporal distances, participants perceived WebSearch as slightly easier to use compared to ConvSearch. Conversely, the negative effect sizes for Social Distance and Hypothetical Distance indicate that when these tasks are framed with near social or hypothetical distances, participants perceived ConvSearch as slightly easier to use compared to WebSearch. However, the differences in perceived ease of use between the two tools under near psychological distances were not statistically significant and are negligible in practical terms.

In the context, as shown in Fig.~\ref{fig:EoU} (f)-(h), significant differences emerged in the perceived ease of use at a far psychological distance in the temporal, social, and hypothetical dimensions with a moderate or large effect size. Specifically, for temporal distance, ConvSearch demonstrated a significantly higher ease of use (\(M = 6.18\), \(SD = 0.73\)) compared to WebSearch (\(M = 5.65\), \(SD = 1.09\)), with a moderate effect size (\(t(15) = -2.14\), \(p = 0.050\), Cohen's \(d = -0.53\)). Similarly, in social distance, ConvSearch (\(M = 6.22\), \(SD = 0.74\)) outperformed WebSearch (\(M = 5.26\), \(SD = 1.10\)), with a large effect size (\(W = 2.00 \), \(p = 0.004\), \(r = -0.96\)). Lastly, in the hypothetical distance dimension, ConvSearch (\(M = 6.30\), \(SD = 0.76\)) was rated higher than WebSearch (\(M = 5.40\), \(SD = 1.23\)), also with a large effect size (\(W = 2.00\), \(p = 0.004\), \(r = -0.96\)). 
For spatial distance (Fig.~\ref{fig:EoU} (e)), although ConvSearch (\(M = 6.28\), \(SD = 0.48\)) scored higher than WebSearch (\(M = 6.02\), \(SD = 0.73\)), the difference was not statistically significant, \(W = 8.00\), \(p = 0.163\), \(r = -0.56\). However, the moderate effect size suggests a meaningful trend favoring ConvSearch under far spatial distances. The positive effect sizes observed for far temporal, social, and hypothetical distances indicate that as the psychological distance of tasks increases, participants increasingly perceive ConvSearch as easier to use compared to WebSearch. These findings partially support H1, indicating that as psychological distance increases, ConvSearch tends to be perceived as easier to use compared to WebSearch, particularly in temporal, social, and hypothetical dimensions.

\subsubsection{Psychological Distance Increases Differences in Perceived Information Credibility (H2 Supported)}

\begin{figure*}[htbp]
    \centering
    \includegraphics[width=0.9\textwidth]{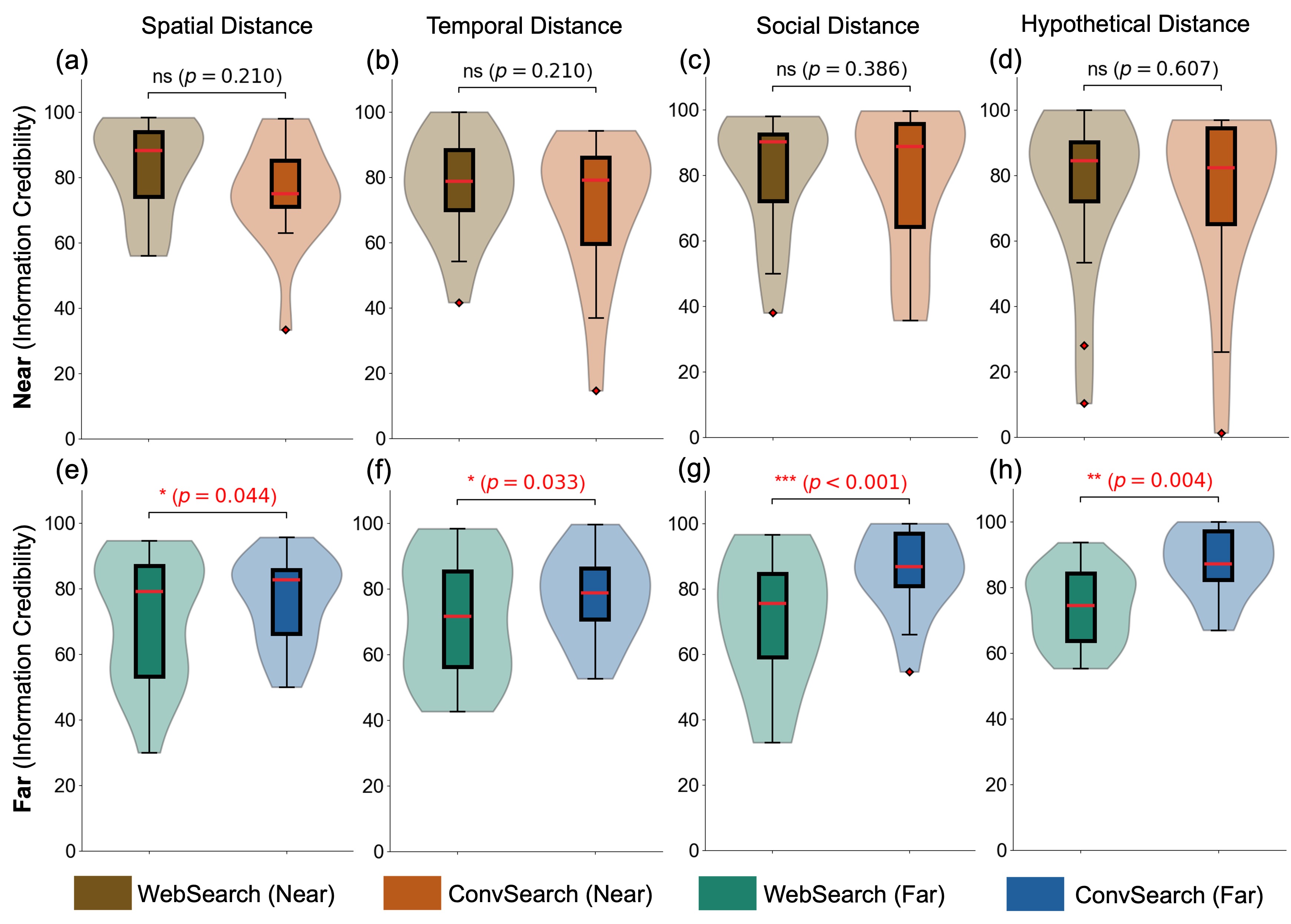}
    \caption{Impact of psychological distance on perceived information credibility for WebSearch and ConvSearch. Panels (a)-(d) illustrate results at a near psychological distance, with significant differences only in spatial distance. Panels (e)-(h) demonstrate increased credibility for ConvSearch at a far psychological distance across all dimensions except temporal distance. Significance is marked as p > 0.05 (ns), p <= 0.05 (*), p < 0.01 (**), or p < 0.001 (***).}
    \Description{This figure presents eight violin plots arranged in two rows and four columns, comparing the perceived information credibility for two search technologies, WebSearch and ConvSearch, across four dimensions of psychological distance: spatial, temporal, social, and hypothetical. Each plot features a pair of violin plots side by side, with the left representing WebSearch and the right representing ConvSearch. The violin plots illustrate the distribution of responses, with a central box plot highlighting the median (red line), the interquartile range (colored box), and outliers (diamond shapes). Statistical significance is indicated with p-values and asterisks from the paired samples statistical tests. In the first row, panels (a) through (d) depict the perceived information credibility at a near psychological distance. The violin plots are colored in shades of beige and brown, representing WebSearch and ConvSearch respectively. The p-values are as follows: (a) Spatial Distance (p = 0.210), marked as non-significant (ns); (b) Temporal Distance (p = 0.210), marked as non-significant (ns); (c) Social Distance (p = 0.386), marked as non-significant (ns); (d) Hypothetical Distance (p = 0.607), marked as non-significant (ns). In the second row, panels (e) through (h) illustrate the perceived information credibility at a far psychological distance. The violin plots are colored in shades of green and blue, again representing WebSearch and ConvSearch respectively. The p-values indicate significant differences in panel (e) Spatial Distance (p = 0.044), marked with one asterisk (*); panels (f) Temporal Distance (p = 0.033), marked with one asterisk (*); (g) Social Distance (p < 0.001), marked with three asterisks (***); (h) Hypothetical Distance (p = 0.004), marked with two asterisks (**).}
    \label{fig:Credibility}
\end{figure*}

At a near psychological distance, as depicted in Fig.~\ref{fig:Credibility} (a)-(d), participants reported similar perceptions of information credibility related to travel between WebSearch (Spatial Distance: \(M = 83.02\), \(SD = 14.47\); Temporal Distance: \(M = 76.88\), \(SD = 15.86\); Social Distance: \(M = 80.69\), \(SD = 18.37\); Hypothetical Distance: \(M = 75.40\), \(SD = 25.22\)) and ConvSearch (Spatial Distance: \(M = 75.46\), \(SD = 15.63\); Temporal Distance: \(M = 70.81\), \(SD = 21.86\); Social Distance: \(M = 78.88\), \(SD = 22.58\); Hypothetical Distance: \(M = 74.40\), \(SD = 27.80\)). Paired samples statistical tests showed non-significant differences with small effect sizes across dimensions (Spatial Distance: \(t(15) = 1.62\), \(p = 0.210\), Cohen's \(d = 0.41\); Temporal Distance: \(t(15) = 1.60\), \(p = 0.210\), Cohen's \(d = 0.40\); Social Distance: \(W = 42.00\), \(p = 0.386\), \(r = -0.30\); Hypothetical Distance: \(t(15) = 0.61\), \(p = 0.607\), Cohen's \(d = 0.15\)). 
The small effect sizes observed at near distances indicate that psychological proximity of the task does not strongly influence users' perceived credibility between the two search modalities. Notably, the slightly positive effect sizes for spatial, temporal, and hypothetical distances suggest a marginal preference for WebSearch in terms of credibility at near distances. Conversely, the negative effect size for social distance implies a slight preference for ConvSearch in this context.

In the context of greater psychological distance, the perceived credibility of information from ConvSearch was significantly higher compared to WebSearch across all measured dimensions, as shown in Fig.~\ref{fig:Credibility} (e)-(h). For spatial distance, ConvSearch was rated as more credible (\(M = 76.38\), \(SD = 13.28\)) than WebSearch (\(M = 70.67\), \(SD = 19.50\)), yielding a significant difference with a moderate effect size, \(t(15) = -2.23\), \(p = 0.044\), Cohen's \(d = -0.56\). In the temporal distance dimension, ConvSearch also demonstrated significantly higher credibility (\(M = 77.94\), \(SD = 13.36\)) compared to WebSearch (\(M = 71.21\), \(SD = 18.73\)), with a moderate effect size, \(t(15) = -2.50\), \(p = 0.033\), Cohen's \(d = -0.62\). Similarly, in the social distance dimension, participants rated the credibility of ConvSearch (\(M = 85.19\), \(SD = 13.11\)) significantly higher than that of WebSearch (\(M = 70.31\), \(SD = 20.38\)), with a large effect size (\(W = 0.00\), \(p < 0.001\), \(r = -1.00\)). Lastly, for hypothetical distance, ConvSearch (\(M = 87.54\), \(SD = 10.11\)) was perceived as more credible than WebSearch (\(M = 74.23\), \(SD = 11.37\)), with a large effect size (\(t(15) = -4.12\), \(p = 0.004\), Cohen's \(d = -1.03\)). 
The moderate and large effect sizes observed at far psychological distances suggest that as the psychological distance of a task increases, ConvSearch becomes substantially more credible in the eyes of users compared to WebSearch. The negative direction of the effect sizes consistently indicates that participants perceived ConvSearch as more credible than WebSearch across all far-distance dimensions.
These findings support H2. They suggest that as psychological distance increases, so does the perceived reliability of information from ConvSearch compared to WebSearch, even when the IR technologies remain unchanged. This indicates that psychological distance can influence user perceptions of information credibility in different contexts.

\subsubsection{Psychological Distance Increases Differences in Perceived Usefulness (H3 Supported)}

\begin{figure*}[htbp]
    \centering
    \includegraphics[width=0.9\textwidth]{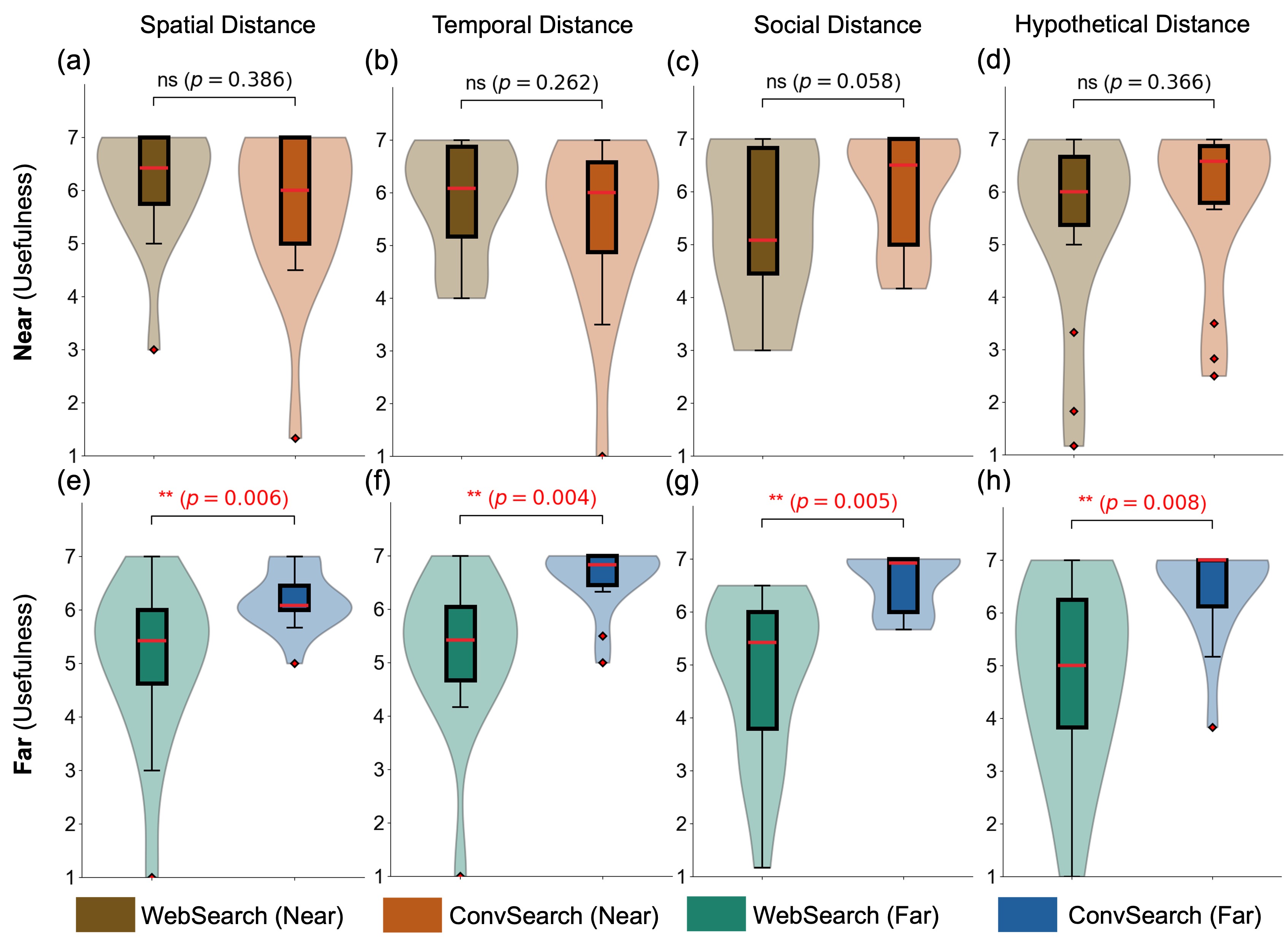}
    \caption{Comparative analysis of perceived usefulness between WebSearch and ConvSearch at varying psychological distances. Panels (a)-(d) show minimal differences at a near psychological distance, except for social distance. Panels (e)-(h) depict significant increases in perceived usefulness for ConvSearch at a far psychological distance across all dimensions. Significance is marked as p > 0.05 (ns), p <= 0.05 (*), p < 0.01 (**), or p < 0.001 (***).}
    \Description{This figure presents eight violin plots arranged in two rows and four columns, comparing the perceived usefulness for two search technologies, WebSearch and ConvSearch, across four dimensions of psychological distance: spatial, temporal, social, and hypothetical. Each plot features a pair of violin plots side by side, with the left representing WebSearch and the right representing ConvSearch. The violin plots illustrate the distribution of responses, with a central box plot highlighting the median (red line), the interquartile range (colored box), and outliers (diamond shapes). Statistical significance is indicated with p-values and asterisks from the paired samples statistical tests. In the first row, panels (a) through (d) depict the perceived usefulness at a near psychological distance. The violin plots are colored in shades of beige and brown, representing WebSearch and ConvSearch respectively. The p-values are as follows: (a) Spatial Distance (p = 0.386), marked as non-significant (ns); (b) Temporal Distance (p = 0.262), marked as non-significant (ns); (c) Social Distance (p = 0.058), marked as non-significant (ns); (d) Hypothetical Distance (p = 0.366), marked as non-significant (ns). In the second row, panels (e) through (h) illustrate the perceived usefulness at a far psychological distance. The violin plots are colored in shades of green and blue, again representing WebSearch and ConvSearch respectively. The p-values indicate significant differences in panel (e) Spatial Distance (p = 0.006), marked with two asterisks (**); panels (f) Temporal Distance (p = 0.004), marked with two asterisks (**); (g) Social Distance (p = 0.005), marked with two asterisks (**); (h) Hypothetical Distance (p = 0.008), marked with two asterisks (**).}
    \label{fig:Usefulness}
\end{figure*}

At a near psychological distance (as illustrated in Fig.~\ref{fig:Usefulness} (a)-(c)), the differences in perceived usefulness between the WebSearch (Spatial Distance: \(M = 6.14\), \(SD = 1.08\); Temporal Distance: \(M = 5.87\), \(SD = 1.09\); Social Distance: \(M = 5.33\), \(SD = 1.50\); Hypothetical Distance: \(M = 5.44\), \(SD = 1.79\)) and ConvSearch (Spatial Distance: \(M = 5.71\), \(SD = 1.50\); Temporal Distance: \(M = 5.47\), \(SD = 1.60\); Social Distance: \(M = 6.04\), \(SD = 1.08\); Hypothetical Distance: \(M = 5.90\), \(SD = 1.48\)) were not significant (Spatial Distance: \(W = 52.00\), \(p = 0.386\), \(r = 0.33\); Temporal Distance: \(t(15) = 1.41\), \(p = 0.262\), Cohen's \(d = 0.35\); Social Distance: \(t(15) = -2.38\), \(p = 0.058\), Cohen's \(d = -0.60\); Hypothetical Distance: \(W = 35.00\), \(p = 0.366\), \(r = -0.33\)).
The small to moderate effect sizes observed at near distances suggest that the psychological proximity of the task does not strongly influence perceived usefulness between the two search modalities. The positive effect sizes for spatial and temporal distances indicate a slight preference for WebSearch in terms of usefulness, whereas the negative effect sizes for social and hypothetical distances suggest a marginal preference for ConvSearch. Notably, the moderate negative effect size for social distance suggests a trend where participants may begin to perceive ConvSearch as slightly more useful than WebSearch, even at near distances.

In contrast, at a far psychological distance, the preference for ConvSearch showed a significant increase. As depicted in Fig.~\ref{fig:Usefulness} (e)-(h), substantial differences emerged in favor of ConvSearch across all dimensions. In spatial distance, ConvSearch (\(M = 6.19\), \(SD = 0.55\)) was perceived as more useful compared to WebSearch (\(M = 5.04\), \(SD = 1.48\)), \(W = 2.00\), \(p = 0.006\), \(r = -0.95\). Similar phenomena were observed in the temporal (ConvSearch: \(M = 6.60\), \(SD = 0.59\); WebSearch: \(M = 5.26\), \(SD = 1.41\); \(W = 3.00\), \(p = 0.004\), \(r = -0.95\)), social (ConvSearch: \(M = 6.53\), \(SD = 0.56\); WebSearch: \(M = 4.85\), \(SD = 1.57\); \(t(15) = -3.66\), \(p = 0.005\), Cohen's \(d = -0.92\)), and hypothetical distances (ConvSearch: \(M = 6.46\), \(SD = 0.90\); WebSearch: \(M = 4.81\), \(SD = 1.78\); \(t(15) = -3.28\), \(p = 0.008\), Cohen's \(d = -0.82\)). 
The large effect sizes observed at four psychological distances indicate that as the psychological distance of a task increases, participants perceive ConvSearch as significantly more useful than WebSearch. The consistent negative direction of the effect sizes suggests a clear and robust preference for ConvSearch in terms of usefulness under conditions of far psychological distance. Notably, the effect size increased from a moderate level at near distances to a large level at far distances, highlighting the particularly strong influence of psychological distance on perceived usefulness.
These findings support H3, demonstrating that as psychological distance increases, so does the perceived usefulness of ConvSearch compared to WebSearch.

\subsubsection{Psychological Distance Increases Differences in Perceived Enjoyment (H4 Supported)}

\begin{figure*}[htbp]
\centering
\includegraphics[width=0.9\textwidth]{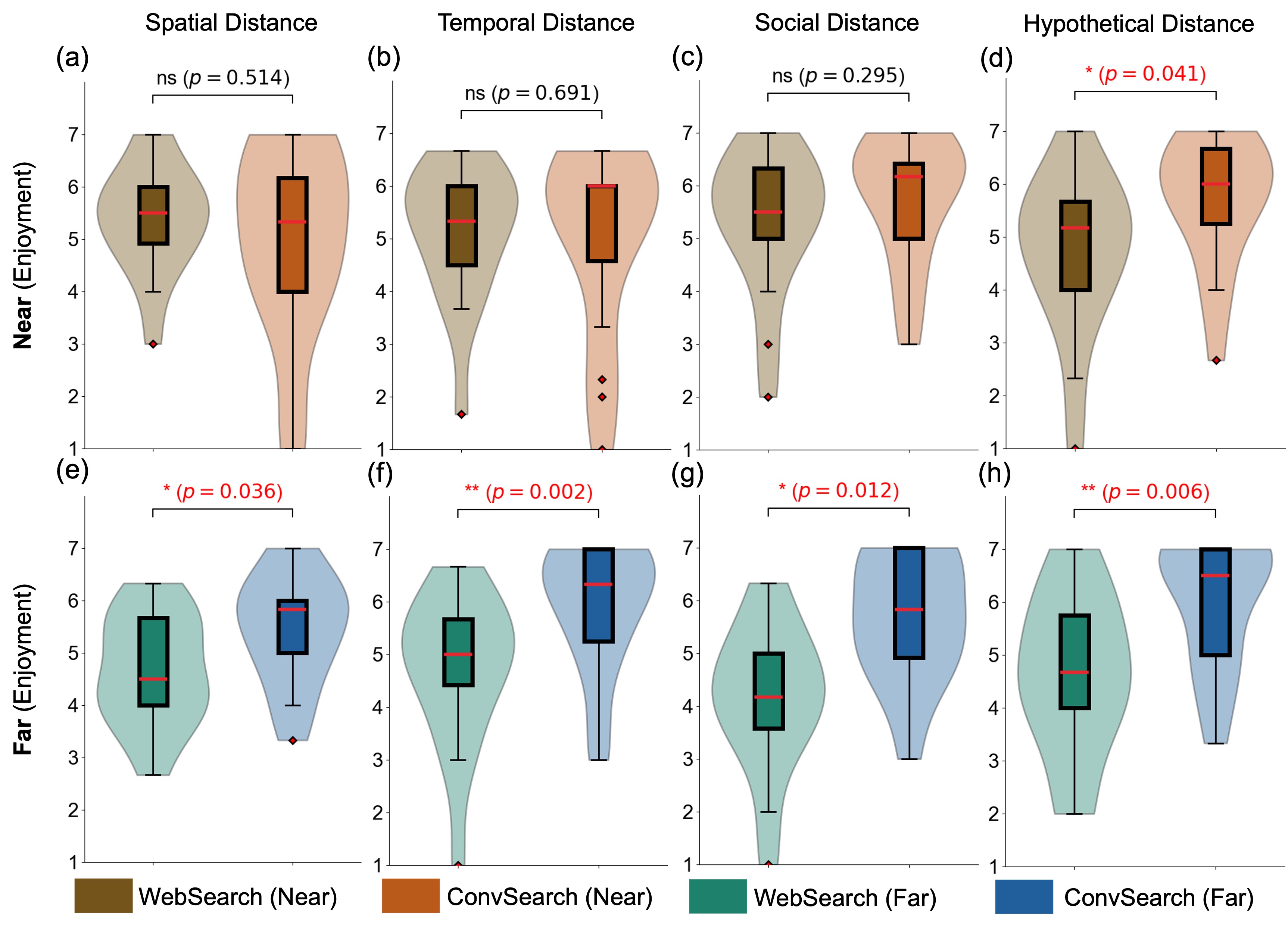}
\caption{Comparative analysis of perceived enjoyment between WebSearch and ConvSearch across four psychological distances. Panels (a)-(c) indicate no significant differences at a near psychological distance, while panel (d) shows a marginal preference for ConvSearch. Panels (e)-(h) demonstrate significant increases in perceived enjoyment for ConvSearch at a far psychological distance across all dimensions. Significance is marked as p > 0.05 (ns), p <= 0.05 (*), p < 0.01 (**), or p < 0.001 (***).}
\Description{This figure presents eight violin plots arranged in two rows and four columns, comparing the perceived enjoyment for two search technologies, WebSearch and ConvSearch, across four dimensions of psychological distance: spatial, temporal, social, and hypothetical. Each plot features a pair of violin plots side by side, with the left representing WebSearch and the right representing ConvSearch. The violin plots illustrate the distribution of responses, with a central box plot highlighting the median (red line), the interquartile range (colored box), and outliers (diamond shapes). Statistical significance is indicated with p-values and asterisks from the paired samples statistical tests. In the first row, panels (a) through (d) depict the perceived enjoyment at a near psychological distance. The violin plots are colored in shades of beige and brown, representing WebSearch and ConvSearch respectively. The p-values are as follows: (a) Spatial Distance (p = 0.514), marked as non-significant (ns); (b) Temporal Distance (p = 0.691), marked as non-significant (ns); (c) Social Distance (p = 0.295), marked as non-significant (ns); (d) Hypothetical Distance (p = 0.041), marked with one asterisk (*). In the second row, panels (e) through (h) illustrate the perceived enjoyment at a far psychological distance. The violin plots are colored in shades of green and blue, again representing WebSearch and ConvSearch respectively. The p-values indicate significant differences in panel (e) Spatial Distance (p = 0.036), marked with one asterisk (*); panels (f) Temporal Distance (p = 0.002), marked with two asterisks (**); (g) Social Distance (p = 0.012), marked with one asterisk (*); (h) Hypothetical Distance (p = 0.006), marked with two asterisks (**).}
\label{fig:Enjoyment}
\end{figure*}

The analysis of perceived enjoyment revealed significant differences between WebSearch and ConvSearch, particularly as psychological distance increased. At a near psychological distance (as shown in Fig.~\ref{fig:Enjoyment} (a)-(c)), across spatial (WebSearch: \(M = 5.40\), \(SD = 1.03\); ConvSearch: \(M = 5.00\), \(SD = 1.81\)), temporal (WebSearch: \(M = 5.13\), \(SD = 1.28\); ConvSearch: \(M = 4.98\), \(SD = 1.78\)), and social dimensions (WebSearch: \(M = 5.33\), \(SD = 1.34\); ConvSearch: \(M = 5.75\), \(SD = 1.21\)), there were no significant differences in perceived enjoyment (Spatial Distance: \(t(15) = 0.78\), \(p = 0.514\), Cohen's \(d = 0.19\); Temporal Distance: \(t(15) = 0.44\), \(p = 0.691\), Cohen's \(d = 0.11\); Social Distance: \(t(15) = -1.30\), \(p = 0.295\), Cohen's \(d = -0.33\)). However, a significant difference was noted in the hypothetical distance dimension (\(t(15) = -2.65\), \(p = 0.041\), Cohen's \(d = -0.66\)), suggesting a higher perceived enjoyment of ConvSearch (\(M = 5.67\), \(SD = 1.22\)) compared to that of WebSearch (\(M = 4.71\), \(SD = 1.50\)) even at a near psychological distance (Fig.~\ref{fig:Enjoyment} (d)).
The small effect sizes observed in spatial and temporal distances at near psychological distances suggest only minimal differences in perceived enjoyment, with a slight preference for WebSearch. In contrast, the moderate negative effect size for social distance and the larger negative effect size for hypothetical distance indicate a preference for ConvSearch in terms of enjoyment, particularly in the hypothetical dimension. These findings suggest that even at near distances, certain dimensions of psychological distance (e.g., hypothetical) may influence user preference for ConvSearch.

As psychological distance expanded, differences in perceived enjoyment became more pronounced. In the far psychological distance condition, as depicted in Fig.~\ref{fig:Enjoyment} (e)-(h), all dimensions showed significant differences (Spatial Distance: \(t(15) = -2.42\), \(p = 0.036\), Cohen's \(d = -0.60\); Temporal Distance: \(t(15) = -4.73\), \(p = 0.002\), Cohen's \(d = -1.18\); Social Distance: \(t(15) = -3.04\), \(p = 0.012\), Cohen's \(d = -0.76\); Hypothetical Distance: \(W = 0.00\), \(p = 0.006\), \(r = -1.00\)) favoring ConvSearch (Spatial Distance: \(M = 5.46\), \(SD = 1.05\); Temporal Distance: \(M = 5.92\), \(SD = 1.23\); Social Distance: \(M = 5.65\), \(SD = 1.21\); Hypothetical Distance: \(M = 6.00\), \(SD = 1.23\)) over WebSearch (Spatial Distance: \(M = 4.67\), \(SD = 1.05\); Temporal Distance: \(M = 4.73\), \(SD = 1.41\); Social Distance: \(M = 4.17\), \(SD = 1.36\); Hypothetical Distance: \(M = 4.63\), \(SD = 1.49\)). Specifically, the hypothetical distance maintained a higher significant difference than that in near psychological distance.
The moderate to large effect sizes observed at far distances for spatial, temporal, social, and hypothetical indicate that as psychological distance increases, ConvSearch is perceived as significantly more enjoyable than WebSearch. The negative direction of the effect sizes consistently reflects a strong preference for ConvSearch in terms of enjoyment. The particularly large effect size for temporal and hypothetical distance suggests that tasks framed with far temporal scenarios (e.g., events occurring in the distant future) or far hypothetical scenarios (e.g., events are unlikely to occur) may elicit the greatest enjoyment from ConvSearch.
These results underscore that as psychological distance increases, the perceived enjoyment of using ConvSearch significantly exceeds that of WebSearch across various dimensions, supporting H4.

\subsubsection{Psychological Distance Increases Differences in Acceptance of ConvSearch (H5 Supported)}

\begin{figure*}[htbp]
    \centering
    \includegraphics[width=1\textwidth]{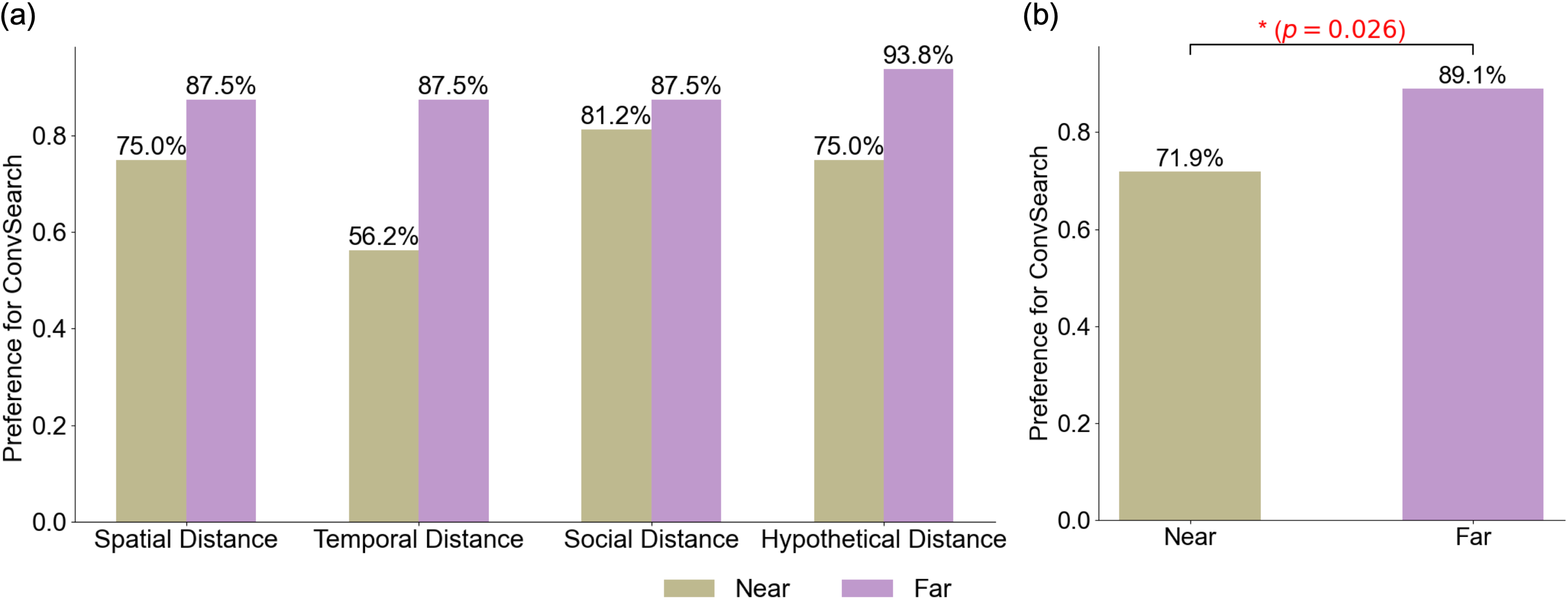}
    \caption{Differential preference for ConvSearch over WebSearch across various psychological distances, shown in two parts: (a) detailed comparison across four dimensions, and (b) aggregated preference at near and far distances. Significance is marked as p > 0.05 (ns), p <= 0.05 (*), p < 0.01 (**), or p < 0.001 (***).}
    \Description{This figure is divided into two parts labeled as (a) and (b), each depicting the preference for ConvSearch over WebSearch across different psychological distances. Part (a) features a horizontal bar chart with eight bars grouped into four pairs, each representing a different psychological dimension: Spatial, Temporal, Social, and Hypothetical. Each pair consists of a beige bar for 'Near' psychological distance and a purple bar for 'Far' psychological distance. The y-axis is scaled from 0 to 1 (0\% to 100\%) and is labeled "Preference for ConvSearch." The exact percentages of preference for ConvSearch are displayed above each bar. For Spatial Distance, the beige bar shows 75.0\% and the purple bar shows 87.5\%. For Temporal Distance, the beige bar is at 56.2\% and the purple bar at 87.5\%. In Social Distance, the beige bar indicates 75.0\% and the purple bar 81.2\%. Finally, for Hypothetical Distance, the beige bar is at 75.0\% and the purple bar significantly higher at 93.8\%. Part (b) of the figure presents a simpler bar chart with just two bars, again in beige and purple, representing the aggregated preference for ConvSearch at Near and Far psychological distances, respectively. The beige bar shows a preference of 71.9\%, while the purple bar shows a higher preference of 89.1\%. The significance of the difference is indicated by an asterisk and the note (p = 0.026) above the bars, referring to the results of a chi-square test.}
    \label{fig:Preference}
\end{figure*}

The impact of psychological distance on the acceptance of ConvSearch supports H5. When the psychological distance is close (as shown in Fig.~\ref{fig:Preference} (a)), the preference for ConvSearch is already evident, with 75.0\% of participants favoring it in the spatial dimension. However, as the distance increases, this preference soars to 87.5\%. Similar patterns are observed in other dimensions. For temporal distance, the preference increases from 56.2\% at a close distance to 87.5\% at a far distance. In social distance, the preference rises from 75.0\% to 81.2\%, and most notably, in hypothetical distance, the preference jumps from 75.0\% to an overwhelming 93.8\%. Upon aggregating the data across all dimensions and analyzing the preference for WebSearch and ConvSearch at near and far psychological distances (Fig.~\ref{fig:Preference} (b)), a chi-square test revealed significant differences ( \(\chi^2(1, N = 128) = 4.97\), \(p = 0.026\)). This result highlights that while participants exhibit a preference for ConvSearch even at a near psychological distance, this preference intensifies as the psychological distance increases. This indicates that participants showed a significant preference for ConvSearch over WebSearch in scenarios characterized by greater psychological distance.

\subsection{Thematic Analysis of User Preferences}

\subsubsection{Unique Themes for Preferring ConvSearch}

The unique thematic of ConvSearch response highlights its distinct advantages over WebSearch.

\textbf{Low Workload}. Many participants (n = 58) reported that using ConvSearch required less effort compared to WebSearch. 
They felt that ConvSearch simplified the information gathering process by eliminating the need to visit multiple websites and compile information independently. Thus, reduced workload was one of the reasons for their preference for ConvSearch. As one participant explained: ``\textit{It was easier to find out what I needed to know and less work on my part to pull it all together.}'' (\textit{P6})

\textbf{Speed and Efficiency}. Participants (n = 43) perceived ConvSearch as faster and more efficient in making travel plans. 
The efficiency they focused on was whether ConvSearch could quickly provide information while meeting their needs to complete travel plans. As one stated, ``\textit{I would choose the chatbot because it saves me time and is more efficient at making a full-bodied itinerary.}'' (\textit{P108})

\textbf{Specific Information Consolidation}. Another major reason participants (n = 43) preferred ConvSearch was its ability to quickly consolidate information from various sources for specific queries. One participant called it, ``\textit{The results the chatbot provides is neater compared to the websearch's results which are just links while the chatbot makes sure to summarize any relevant info and provides it to me.}'' (\textit{P96})

\textbf{Interactive and Conversational}. ConvSearch's natural dialogue offers participants (n = 18) a superior interactive experience. Some participants described the interaction as akin to conversing with a human. For instance: ``\textit{I would love to use the chatbot since it feels like having a conversation with a human and getting personalized advice from them. This makes the information easier to understand and apply to my travel plans.}'' (\textit{P77})

\subsubsection{Unique Themes for Preferring WebSearch}

WebSearch has fewer unique themes than ConvSearch, but it demonstrates the irreplaceable advantages of WebSearch.

\textbf{Visual Confirmation and Source Credibility}. Some participants (n = 12) mentioned that WebSearch provides information not only in text but also in images. In contrast to ConvSearch, WebSearch enables participants to identify the sources of information. This capability allows participants to evaluate the reliability of the information. The ability to visually confirm and select information sources plays a significant role in users' decision-making processes. One participant called it, ``\textit{I like to see what I am getting into and have a visual to confirm that I am making a good choice. It makes me feel more at ease with my decision.}'' (\textit{P2})

Specifically, for time-sensitive information, WebSearch is uniquely able to help users verify timeliness. As one participant commented: ``\textit{I'd rather use the web search so I can see the sources of where I'm getting my information from. Also can see how recent the information is.}'' (\textit{P121})

\textbf{Habit and Experience}. A minority of participants (n = 2) prefer WebSearch due to habitual use and greater experience with it. One participant called it, ``\textit{I think that I can access more information and feel more confident about where that information is coming from on a web search. I've been doing web searches forever and feel confident in my ability to find the best and most credible information through them and I can bookmark my findings for later.}'' (\textit{P10})

\subsubsection{Themes Found in Both Systems}

Some themes emerged in participants' reasons for preferring both ConvSearch and WebSearch. These themes generally reflect subjective perceptions rather than functional differences between ConvSearch and WebSearch, indicating personalized preferences.

\textbf{Reliability and Accuracy}. Some participants (n = 7) cited the inaccuracy and unreliability of ConvSearch as reasons for preferring WebSearch. As one participant explained: ``\textit{I would use the web search, although it may be more cumbersome. The reason is that I am able to see the direct sources for the information I receive, unlike the chatbot. The chatbot was easier, just not reliable enough for me.}'' (\textit{P80})

Some participants (n = 12) expressed contrary views, believing that web search results might contain more misinformation, making conversational search more reliable and accurate. One noted: ``\textit{I would choose the chatbot rather than a web search because I would not have to look through multiple web pages for information and potentially run into misinformation and instead I can get trustworthy answers from the chatbot itself.}'' (\textit{P117})

\textbf{Discovery and Exploration}. WebSearch's result list offers participants (n = 4) greater freedom to explore information and gain new ideas and interests during the exploration process. One participant said: ``\textit{I would rather use the web search because I like to look things up and decide which sites I want to use for information on my own. Also, when you do a web search you may find things that spark new ideas and interests. I think it might be useful to use a chatbot to get quick bits of information, and the use the web to search for the things you are most interested in.}'' (\textit{P18})

For some participants (n = 9), the more comprehensive information provided by ConvSearch serves as a foundation to gain new ideas and further explore the information. One participant said: ``\textit{I think it's a good foundational point to work off of, especially when creating an itinerary at a location I know nothing about. Chatbot gave me ideas of what to explore further and potentially search to see reviews before finalizing.}'' (\textit{P116})

\textbf{Combining Two Tools \footnote{Detailed data can be found in the Appendix (A.3), uploaded with Supplementary Material.}}. Some participants (n = 9), despite being explicitly instructed to provide reasons for their preference for either web search or conversational search, still suggested combining both tools. As one noted: ``\textit{Chatbot makes it easier to get a lot of info as a great starting point. I would then use a search engine to verify the info from the chatbot.}'' (\textit{P32})

\textbf{Comprehensive Information}. For planning trips, one participant prefers to refer to real experiences shared by other travelers. WebSearch can provide these authentic experiences, which ConvSearch cannot. Therefore, WebSearch offers more comprehensive information. ``\textit{I would prefer a search box because it gives more details and articles from previous experiences.}'' (\textit{P43})

Participants (n = 7) who believed ConvSearch provided more information did so because it delivered more content than WebSearch for the same amount of effort or time. For instance: ``\textit{I think the chatbot could give me more information depending on the prompt I gave it.}'' (\textit{P4})
\section{Discussion}

Our findings highlight the important role of psychological distance of information seeking tasks in shaping user perceptions and preferences toward different IR technologies. As psychological distance increased, participants showed a preference for LLM-powered conversational search over web search, under fixed condition ordering where web search preceded conversational search. They perceived conversational search as easier to use, more credible, useful, and enjoyable. Consequently, user acceptance of LLM-powered conversational search also rose with increasing psychological distance. These valuable insights highlight how the ubiquitous factor of psychological distance informs design considerations for the nascent technology that is generative AI-based conversational search.

\subsection{Deep Impact of Psychological Distance in Information Seeking Tasks (RQ)}

Our findings support CLT's suggested existence of a relationship between psychological distance and user evaluations, and hence choice of IR technologies \cite{liberman2014traversing, trope2010construal, trope2007construal}. People's choices and preferences are based on their understanding of events, which is in turn affected by psychological distance \cite{trope2010construal}. Our research provides a comprehensive view from the nature of tasks to user preferences, not only enhancing our understanding of the effects of task psychological distance but also offering practical guidance for future IR system design. By addressing user information needs at different psychological distances, IR systems can further improve user interaction experiences and satisfaction.

\subsubsection{Greater Psychological Distance Makes Conversational Search Easier to Use (H1)}
\label{subsubsec:discussionh1}

Our results suggest that with tasks with greater psychological distance, participants find conversational search easier to use than web search. Given that far-distance users tend to think in more abstract and general concepts \cite{trope2007construal}, this preference is likely explained by conversational search better aligning with information needs for these participants. LLMs have been shown to be adaptable to the construal level required by a user \cite{yoo2023large}, and can provide integrated and summarized information tailored to specific questions \cite{kaushik2023comparing}. Search engines, naturally, lack this capability, and require users to manually filter and summarize information from a list of results. Far-distance tasks that require high-level and abstract information are therefore more easily accomplished with LLM-powered search.

Our thematic analysis also supports this perspective. Participants cited \textit{Low Workload} and \textit{Specific Information Consolidation} as reasons for preferring conversational search. The \textit{Low Workload} theme is directly linked to perceived ease of use, as less workload allows users to meet their needs more easily. \textit{Specific Information Consolidation} can also be linked to \textit{Low Workload} in this context, illustrating how conversational search's ability to integrate information underpins its ease of use.

\subsubsection{Greater Psychological Distance Makes Conversational Search More Credible (H2)}
\label{subsubsec:discussionh2}

Our results show that tasks with greater psychological distance lead participants to perceive higher information credibility in LLM-powered conversational search than web search. This perception is further elucidated by our thematic analysis. In the themes reported by participants, \textit{Reliability and Accuracy} are closely related to their perception of information credibility. Participants who prefer conversational search believe it provides more accurate answers by helping them filter results. In contrast, web search results often provide too many options, which sometimes do not align with the information sought by users.

Further analysis of the three dimensions of credibility\footnote{Specific data are presented in the Appendix (A.4), uploaded with Supplementary Material.} reveals that although user perceptions of the accuracy and believability of conversational search tend to improve with greater psychological distance, the perceived enhancement in information authenticity is less pronounced. Unlike accuracy and believability, authenticity is related more to the legitimacy of the source of a message, rather than the factuality of the message content itself \cite{kang2010measuring}. In the context of our travel task, authentic content might come from, for instance, a passionate and well-liked travel blogger. LLMs presently still struggle to reliably provide source attributions \cite{maynez2020faithfulness, bender2021dangers}, and themselves are unlikely to be regarded as "authentic" as real human writers; this may account for the fact that they score relatively lower on this metric. We note, however, that this may vary with the nature of a given task: certain tasks, particularly those involving searching for lived human experiences, may emphasize perceived authenticity more than other tasks relying on purely factual information. 

Recent advancements in AI-powered search systems, such as OpenAI's integration of web browsing capabilities \cite{searchgpt}, have begun to address some limitations of standalone conversational or web search tools. These systems now offer source attribution and access to real-time information, significantly enhancing their utility. However, challenges remain in dynamically adapting to different user tasks and mitigating the confirmatory bias inherent in conversational search.

\subsubsection{Greater Psychological Distance Makes Conversational Search More Useful and Enjoyable (H3, H4)}

As psychological distance increases, user perceptions of the usefulness and enjoyment of LLM-powered conversational search improve significantly. 
This underscores that increasing the psychological distance of a task can enhance user perceptions of the usefulness and enjoyment of conversational search, even without changes in the technology itself. This finding aligns with the TAM and its extensions \cite{davis1989perceived, davis1992extrinsic}, which shows the positive interlink between perceived ease of use \cite{adams1992perceived}, credibility \cite{ayeh2015travellers}, enjoyment, and usefulness. In our thematic analysis, \textit{Speed and Efficiency} are directly related to user perception of usefulness. Participants find conversational search useful because it allows them to complete tasks faster and more efficiently than web search. The perception of enjoyment of conversational search results from enhanced interactivity through natural dialogue (\textit{Interactive and Conversational}). Some participants feel that interacting with conversational systems is like conversing with another human, aligning with research findings that humanlike qualities are an influential factor in making chatbots enjoyable \cite{folstad2020users}.

This finding is particularly noteworthy as it suggests a link between two seemingly unconnected topics - psychological distance of a search task and enjoyment using a search system. Psychological distance and enjoyment have been studied in areas such as teaching \cite{bussing2020topic} and sports watching \cite{lim2012getting}, where user enjoyment is naturally paramount to user outcomes. However, the fact that this relation persists even in a task like IR - not something intuitively perceived as inducing a high level of enjoyment - suggests that IR system designers should consider psychological distance as an important factor in determining user satisfaction and outcomes.

\subsection{Overall User Acceptance of Conversational Search (H5)}
\label{subsec:discussionUserAcceptance}

Our findings align with the TAM and its extensions \cite{davis1989perceived, davis1992extrinsic, davis1992extrinsic}, suggesting that improvements in the previous four user metrics culminate in \textit{increased user acceptance of LLM-powered IR systems as psychological distances increase (H5)}. Additionally, we found that controlling for psychological distance, users generally prefer LLM-powered conversational search over traditional web search. 
For common information seeking tasks like travel planning, this preference may stem from the inherent advantages of LLM-powered conversational search in areas such as \textit{Low Workload}, \textit{Speed and Efficiency}, \textit{Specific Information Consolidation}, and \textit{Interactive and Conversational} capabilities. These characteristics enable LLM-powered conversational search systems to more effectively meet users' information needs in daily tasks.

Another important aspect influencing this increased acceptance is the perceived credibility of LLM-powered conversational search systems, particularly when psychological distance is high. Unlike other factors, which directly influence user acceptance, credibility exerts its impact on acceptance indirectly. As discussed in Section \ref{subsubsec:discussionh2}, tasks with greater psychological distance lead to higher perceived information credibility for conversational search compared to web search. For example, some participants reported that conversational search is more reliable, as it filters results and provides accurate responses. This aligns with prior research showing that perceived credibility enhances user trust in automated systems by alleviating concerns about misinformation or irrelevant results \cite{fogg1999elements, kim2008trust}. When users perceive a system as credible, they are more likely to view it as reliable and efficient, fostering greater acceptance. However, our findings also highlight nuances in how credibility influences acceptance. While conversational search excels in perceived accuracy and believability, its perceived authenticity remains lower compared to web search, as LLMs often lack robust source attributions \cite{maynez2020faithfulness, bender2021dangers}. This limitation may partially constrain user acceptance in tasks where source legitimacy is critical, such as those requiring lived human experiences or expert opinions. Nevertheless, for tasks with greater psychological distance—where users prioritize abstract or general information over specific, source-authenticated details—credibility perceptions of conversational search systems are sufficiently high to drive overall acceptance.

\subsection{Design Implications}

Our research has confirmed the impact of psychological distance in information seeking tasks on user perception and preferences. Conversational search, with its ability to provide information from an abstract perspective, better meets user needs when the psychological distance is greater. 
However, current conversational search systems exhibit a fixed level of abstraction in information provision, regardless of human cognitive state \cite{kirshner2024gpt}. This limits the effectiveness of LLMs in providing information, as it does not fully meet the diverse judgment needs inherent in human cognition \cite{santurkar2023whose}. Past HCI research has also touched on similar views when discussing the design of conversational search, as they identified user demands for different levels of information, with some preferring direct access to answers, while others favored receiving broad background information \cite{vtyurina2017exploring}. This aligns with the two information levels that we have elucidated. However, these studies did not identify the primary factors influencing user information needs.

We propose using psychological distance as a consideration to enhance the diversity of information levels provided by conversational search systems. By considering psychological distance, we can not only increase the diversity of information, but also better meet the information needs at the user's current cognitive state. 

\subsubsection{Impact on Conversational Search Design}

To leverage the capabilities of LLM-powered conversational search systems, it is important to assess the psychological distance of information seeking tasks through dialogue content and treat it as a significant attribute. An individual's current psychological distance to a task manifests in the way they describe events or actions. This involves omitting features and details as distance increases, e.g. "having breakfast in a Parisian café by the Riviera" vs. "a romantic vacation in France" \cite{trope2007construal}. Using LLMs or other natural language processing means, IR systems can automatically analyze a user's inputs for either concrete details or abstract concepts, to determine the construal level (and hence psychological distance) a user is operating at. Additionally, systems may also use the capability of conversational search \cite{radlinski2017theoretical} to pose supplementary questions about the user's goals along the four dimensions of spatial, temporal, social, and hypothetical distance, further refining the evaluation of psychological distance. 

Conversational search could adjust its IR strategy based on the psychological distance perceived by users. For tasks with a closer psychological distance, users typically seek specific and actionable information. In such cases, search results should focus on providing detailed steps and practical guidance. For example, when planning a trip, the system could provide specific travel itineraries, recommended accommodations, and dining options. This approach not only satisfies the user's low-level, specific information needs but also aids in making quick decisions and executing plans. 

Conversely, for tasks with greater psychological distance, users tend to seek higher-level, more abstract information. Here, conversational search could provide an overview of the topic, helping users quickly grasp the core content and related broad information. For instance, if a user is considering potential future travel destinations, the system could provide summaries of various cities, their cultural highlights, and culinary recommendations, thus providing a comprehensive understanding and sparking further exploration interest.

Additionally, to further optimize the performance of LLM-powered conversational search, we also provide recommendations for fine-tuning LLMs specifically for IR. We suggest labeling internet data used for model fine-tuning to distinguish between specific and abstract information in response to queries. Through such fine-tuning, the model can more accurately understand the level of information, thereby providing more personalized and precise search results. Implementing these recommendations, the optimized conversational search will not only better meet user information needs but also enhance the search experience and strengthen user preferences.

\subsubsection{Combine Conversational Search and Web Search}

Currently, clear guidance on integrating conversational search with web search is lacking, often leaving users to choose their preferred tool without knowing which would better satisfy their information needs. 
Considering the psychological distance of tasks, it is advisable to proactively offer tools that more closely align with current information needs based on this criterion. Therefore, we propose several recommendations for systems that integrate conversational and web search.

Initially, it would be beneficial to present a conversational search interface to users. This design leverages the natural conversational abilities of LLMs to reduce the effort users must expend in formulating queries \cite{radlinski2017theoretical}. The system can utilize the LLM's understanding capabilities to assess the psychological distance of the user's current task through dialogue and to ask supplementary questions \cite{radlinski2017theoretical}, thereby more accurately gathering information about the task's psychological dimensions. This step can enable the system to tailor the presentation of information more effectively.

For tasks with closer psychological distances, we recommend returning a list of web search results. 
For example, when searching for local restaurants or immediate weather forecasts, web search can provide timely and specific information that directly meets users' needs.
However, for tasks with a greater psychological distance, we suggest providing synthesized information generated by conversational search. 
For example, when exploring potential travel destinations or learning about new technologies, conversational search can provide a comprehensive information framework to help users build a preliminary understanding of the topic.

Our findings also suggest that credibility perceptions, influenced by psychological distance, can guide how combined IR systems present and contextualize information. For tasks with closer psychological distances, where users may prioritize source authenticity, systems should emphasize source attributions in web search results, enhancing perceived credibility through transparency and trustworthiness. Conversely, for tasks with greater psychological distances, where users value abstract and synthesized information, conversational search could focus on providing high-level overviews while clearly communicating the limitations of LLM-generated content, such as potential biases or incomplete source references.

Although past HCI research did not focus on the psychological distance of tasks, it explored factors influencing user choices between conversational and web search in specific task scenarios. A study on programming tasks noted that programmers opt for conversational search when objectives are clear, seeking results aligned with their goals \cite{yen2024search}. This phenomenon, stemming from the confirmatory nature of conversational search, known as the echo chamber effect, has been extensively studied in HCI \cite{sharma2024generative, tanprasert2024debate}. 

To optimize this, some studies have suggested designs that allow conversational systems to actively present different viewpoints \cite{feltwell2020broadening} and designs that proactively remind users of current information biases in conversations \cite{sharma2024generative}. 
However, we propose that integrating conversational and web search, guided by psychological distance, offers a more holistic solution. By dynamically adapting the balance between the two modes based on user needs, systems can mitigate biases while enhancing the diversity and relevance of presented information. 
Additionally, HCI research discussing user trust in ChatGPT versus web search noted that generative AI's lack of information sources reduces user trust \cite{jung2024we}. Yet, its assistance in ideation encourages its use. Our participants expressed similar views \footnote{Specific data are presented in the Appendix (A.3), uploaded with Supplementary Material.}, preferring to obtain preliminary information through conversational search, then verify this information and acquire more details through web search.

\subsection{Limitations}

This study has several limitations. First, our study only considered information seeking tasks in travel scenarios. These tasks, being general in nature, may have influenced user perceptions of LLM-powered conversational search, which is better suited for general rather than specialized tasks. This limitation might obscure the specific differences in user perceptions between web search and LLM-powered conversational search. For instance, we observed no significant differences in most metrics when the psychological distance was near. This pattern may not apply to other scenarios and also prevented us from further exploring the specific impacts of tasks with near psychological distance on both types of IR tools.
However, with web search as a control, we found no significant differences in most ratings between the two IR tools when the psychological distance was near, but significant differences emerged when the psychological distance was far. This outcome still addresses our main research question and reveals a general principle: Users prefer LLM-powered conversational search when the psychological distance is greater.

Furthermore, our study did not counterbalance the order of conditions, as web search was always presented first, followed by conversational search. This decision was made for two primary reasons. First, conversational search may lead participants to use a wider variety of prompts, resulting in significant variability in the results they obtain. If conversational search were presented first, the knowledge gained during this task could significantly influence the behavior of participants in the subsequent task. In contrast, web search is a more familiar tool, and participants are more likely to exhibit consistent behaviors and obtain relatively similar results during this task. By placing web search first, we aimed to create a pseudo-control condition, ensuring greater consistency across participants when they transitioned to the conversational search condition. Second, any potential bias introduced by the fixed order of conditions would therefore be consistent across all experimental groups, which may not compromise the validity of our primary findings.
That said, the fixed order remains a limitation, as it may have influenced the effectiveness of our psychological distance manipulations. For instance, participants might have adjusted their perceptions of psychological distance after completing the first task (web search), which could have affected their subsequent interactions with conversational search. Future studies could adopt a counterbalanced design to explore how task order interacts with psychological distance manipulations and influences user perceptions and preferences.

Finally, the custom search interface we implemented does not fully represent a real, commercial search page; it lacks features like image previews and related searches, especially compared to the actual Google Search website. However, since the search results are sourced from the genuine Google Search API, we consider them to still fairly represent a real search interaction.
\section{Conclusion}

In this study, we employed a mixed-methods experiment to explore how psychological distance in information seeking tasks affects user perceptions and preferences for IR technologies. We found that user preference for conversational search increases with the psychological distance of the task, and that this preference is primarily driven by users' perceptions of ease of use, credibility, usefulness, and enjoyment. Notably, these perceptions are significantly enhanced in conversational search when the task's psychological distance is greater. By integrating CLT with IR technologies, our study underscores the psychological distance in tasks as a key factor that influences user perceptions and shapes user preferences. We provide design recommendations for conversational systems to adjust their output based on task nature, and for conversation-web search hybrid systems that take into account the nature of different tasks. We acknowledge that the fixed order of conditions may have introduced order effects; while we chose this to reduce variability in participant behavior, future research could adopt a counterbalanced design to validate and extend our findings.


\begin{acks}
This research is supported by the Spotlight Research Project of National Dong Hwa University (NDHU: 114T2560-10), the Yale-NUS College grant (A-8001353-00-00), the NUS IT’s Research Computing group (NUSREC-HPC-00001), the Singapore Ministry of Education Academic Research Fund (A-8002610-00-00), and AWS. We sincerely thank our participants for their participation and anonymous reviewers for their valuable comments and suggestions on this work.
\end{acks}

\bibliographystyle{ACM-Reference-Format}
\bibliography{sample-base}


\end{document}